\documentclass[aps,twocolumn,raggedbottom,prb,showpacs,nobalancelastpage,amssymb,groupedaddress]{revtex4}
\usepackage{graphicx}
\usepackage{amsmath}
\newcommand{\phd}{\phantom{\dagger}}

\begin{document}
\title{Linear conductance of an interacting carbon nanotube ring}
\author{Matthis\, Eroms, }
\affiliation{ Institut f\"{u}r Physikalische Chemie,
Universit\"at Heidelberg, 69120 Heidelberg, Germany}
\author{Leonhard\, Mayrhofer and Milena\, Grifoni}
\affiliation{Institut f\"{u}r Theoretische Physik, Universit\"at
Regensburg, 93035 Regensburg, Germany}
\date{\today}
\begin{abstract}
Linear transport through a single-walled carbon
nanotube ring, pierced by a magnetic field and capacitively coupled to a
gate voltage source, is investigated starting from a model of interacting $p_z$-electrons. The dc-conductance,
calculated in the limit of weak tunneling between the ring and the leads,
displays a periodic resonance pattern determined by the interplay between Coulomb interactions and quantum interference
phenomena. Coulomb blockade effects are manifested in the absence of resonances for any applied flux in some
gate voltage regions; the periodicity as a function of the applied flux can be smaller or larger than a
flux quantum depending on the nanotube band mismatch.
\end{abstract}
%%\subtitle{Do you have a subtitle?\\ If so, write it here}
%\author{Matthis Eroms \and Leonhard Mayrhofer \and Milena Grifoni
% \thanks is optional - remove next line if not needed
%\thanks{\emph{Present address:} Insert the address here if needed}%
                    % Do not remove
%
%\offprints{}          % Insert a name or remove this line
%
%\institute{Theoretische Physik, Universit\"at Regensburg, 93040 Regensburg, Germany}
%
%\date{Received: date / Revised version: date}
% The correct dates will be entered by Springer
%Insert your abstract here.
%
 % end of PACS codes
 %end of abstract
%
\maketitle
\section{Introduction}
\label{intro}
Mesoscopic rings threaded by a magnetic field represent an
important tool for the investigation of quantum interference
phenomena.  The archetype example is the well known Aharonov-Bohm
effect \cite{Aharonov59}, where the conductance of a clean ring
exhibits a periodicity of one flux quantum $\Phi_0=h/e$. However,
impurities \cite{Altshuler81} and interactions
%\cite{Loss93,
\cite{Jagla93} can change this periodicity. In
particular when transport through a one-dimensional ring is
considered, interactions lead to spin-charge separation, such that
 the most important contribution to the conductance arises when both charge and spin
excitations propagate from one contact to the other arriving at
the drain at the same time \cite{Jagla93}.
 Additionally, the dc-conductance of an interacting one-dimensional ring shows
 Coulomb oscillations with peak positions depending on the applied magnetic field
 and interaction strength \cite{Kinaret98,Pletyukhov06}.

Among quasi-one dimensional systems, single-walled carbon
nanotubes \cite{Saito98} (SWNTs) have been proved to be extremely
interesting to probe electron-electron correlation effects.
Luttinger liquid behavior, leading to power-law dependence of
various quantities, has been predicted theoretically
\cite{Egger97,Kane97} and observed experimentally \cite{Bockrath99,Tans97,Postma01,Lee04}
in long, \emph{straight} nanotubes.

Moreover, as typical of low-dimensional systems, short, \emph{straight} carbon nanotubes weakly attached to
leads exhibit Coulomb blockade at low temperatures \cite{Tans97}.
In metallic SWNTs two bands
cross at the Fermi energy. Together with the spin degree
this leads to the formation of electron shells,
each accommodating up to four electrons.
As a result, a characteristic even-odd \cite{Cobden02}  or fourfold
 \cite{Liang02,Sapmaz05,Moriyama05} periodicity of the Coulomb diamond size as a
function of the gate voltage is found. Recently, spin-orbit effects
in carbon nanotube quantum dots have been observed as well \cite{Kuemmeth08}. While the Coulomb
blockade can be explained merely by the ground state
properties of a SWNT, the determination of the current
at higher bias voltages requires the inclusion of transitions
of the system to electronic excitations. In \cite{Oreg00}
a mean-field treatment of the electron-electron interactions has been
invoked to calculate the energy spectrum, while in \cite{Mayrhofer06} a
bosonization approach, valid for nanotubes with
moderate-to-large diameters ($\emptyset>1.5$nm), has been used.
Within the bosonization approach the fermionic ground state as
well as the fermionic and bosonic excitations can be calculated.
For small diameter nanotubes, short range interactions lead to pronounced
exchange effects \cite{Oreg00,Mayrhofer08} which result in
an experimentally detectable \cite{Moriyama05} singlet-triplet splitting.

\begin{figure}
        \includegraphics[width=\columnwidth,angle=0]{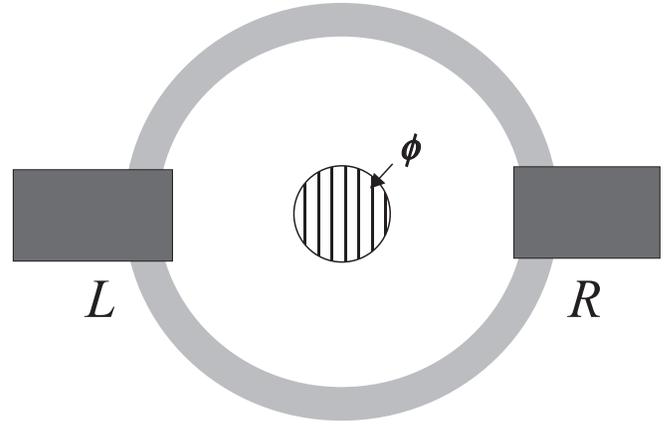}
    \caption{Top view of a SWNT ring contacted to extended left and right leads and threaded by a magnetic flux $\phi$.
    The ring is also
    capacitively  coupled to a  gate voltage $V_g$ plane situated beneath the ring (not shown). }
    \label{ring}
\end{figure}
Generically, nanotubes have linear or curved shape. Individual circular
single-wall carbon nanotubes have been observed in \cite{Liu97,Martel99,Shea00}.
However, these systems have been poorly experimentally investigated so far.
Also from the theoretical point of view, only few works address the properties of
\emph{toroidal} nanotubes \cite{Lin98,Odinstov99,Rollbühler99,Latil03,Rocha04,Zaho04,Jack07}.
Moreover, except for the studies \cite{Odinstov99,Rollbühler99},
on the persistent current and on the conductance of toroidal SWNTs, respectively, all the
remaining theoretical works  neglect electron-electron
interactions effects. However, as for the case of straight SWNTs discussed above,
electron correlation effects are expected to crucially influence the energy spectrum
and transport properties of toroidal SWNTs.  Indeed, in \cite{Odinstov99} it is found
that for interacting SWNTs rings the persistent current pattern corresponds to the constant interaction model,
with a fine structure stemming from exchange correlations.
Results for the conductance of a SWNT ring weakly contacted to rings have been presented so far only in the short report
\cite{Rollbühler99}, where a conductance resonance pattern depending on the interaction strength is reported.
Moreover, a detailed derivation of the conductance formula for SWNT rings is missing in the short report \cite{Rollbühler99}.
In this work we generalize the analysis of \cite{Kinaret98} on the conductance of interacting spinless electrons in
a one-dimensional  ring to the case of a three-dimensional toroidal, metallic SWNT at low energies. To this extent we start from a model of interacting $p_z$ electrons on a graphene lattice at low energies and impose
 periodic boundary conditions along the nanotube circumference and twisting boundary conditions along the tube axis. At low energies only the
lowest transverse energy sub-bands of the ring contribute to transport, so that the problem becomes effectively one-dimensional in momentum space. The three-dimensional shape of the SWNT orbitals in real space, however, crucially determines the final conductance formula.  Indeed,
 in contrast to \cite{Rollbühler99}, we find the
absence of interference terms  between anti-clockwise and clockwise circulating electrons in the conductance formula.
 This is due to the
  localized character of the $p_z$ orbitals and the fact that for a realistic nanotube the contacts are extended.
 Additionally, we predict an eight electron periodicity of the conductance resonance pattern as a function of the
 applied gate voltage. Coulomb blockade effects are clearly visible in that for some gate voltage ranges a resonance condition is not met for any value of the applied flux.
 The resonance pattern is also periodic as a function of the applied magnetic field,
 with a periodicity which can be larger or smaller than one flux quantum for nanotubes with a band mismatch.

 The manuscript is organized as follows. The total Hamiltonian and its low energy spectrum are discussed in Secs. \ref{Sec1:model ham} and \ref{Sec2:low energy}. Specifically, the low energy Hamiltonian of the non-interacting system possesses two linear branches, corresponding to clockwise and anti-clockwise motion, crossing at the two non-equivalent Fermi points
of the graphene lattice. By inclusion of the dominant forward scattering processes only, the interacting Hamiltonian is diagonalized exactly by standard bosonization techniques \cite{Delft98}.
In \ref{Sec3:linear transport} the conductance formula is derived, while Sec. \ref{Sec4:Greens funct} is dedicated to the evaluation of the
Green's functions of the interacting SWNT ring. Finally the conductance resonance pattern is discussed in Sec. \ref{Sec5:results cond}, where also conclusions are drawn.

%\section{Carbon nanotube ring system} \label{cnr system}
\section{The total Hamiltonian} \label{Sec1:model ham}
We consider a ring made of a toroidal metallic single-wall carbon
nanotube  coupled to a source and drain electrode via
tunneling contacts, cf. figure \ref{ring}. The torus is
capacitively connected to a gate electrode with gate voltage
$V_g$, which changes the chemical potential of the ring. Further,
a magnetic flux $\phi$ threads the center of the torus. We wish to
study how the two parameters $V_g$ and $\phi$ influence the linear
conductance of the system.
The model Hamiltonian reads
\begin{eqnarray}
\hat{H} = \hat{H}_{\mathrm{ring}} + \hat{H}_L + \hat{H}_R +
\hat{H}_T + \hat{H}_{ext},  \label{total H}
\end{eqnarray}
where $\hat{H}_{\mathrm{ring}}$, which also includes the
parameters $V_g$ and $\phi$, describes the physics of the isolated
SWNT ring and will be discussed in the next section. The second
and third term refer to the metallic left  and right contacts,
described here  as Fermi gases of non-interacting electrons. They
read ($\alpha=L/R$)
%\begin{eqnarray}
$\hat{H}_{\alpha} = \sum_{{\vec q} \sigma} \varepsilon_\alpha(q)
c^\dagger_{\alpha {\vec q} \sigma} c^{\phd}_{\alpha {\vec q}
\sigma}$,
%\end{eqnarray}
where $\varepsilon_\alpha(q)$ is the energy dispersion relation of
the lead $\alpha$ and $c_{\alpha {\vec q} \sigma}$ is an operator
annihilating an electron with wave vector ${\vec q}$ and spin
$\sigma$.
\\
The  term $\hat{H}_T$ is the Hamiltonian describing tunneling
between the ring and the leads. Therefore it consists of two
terms, $\hat{H}_T=\hat{H}_{TL} + \hat{H}_{TR}$, and reads
\begin{equation}
\hat H_{T} = \sum_{\alpha=R,L}\sum_\sigma \int d^3 r
(T_{\alpha}(\vec r) \Psi_\sigma ^\dagger (\vec r) \Phi_{\sigma
\alpha}(\vec r) + h.c.),
\end{equation}
where $\Psi_\sigma^\dagger ({\vec r})$ and
$\Phi_{\sigma\alpha}(\vec r)=\sum_{\vec q}\phi_{\vec q}(\vec
r)c_{{\vec q}\sigma \alpha}$ are electron operators in the
dot and in the leads, respectively, and $T_\alpha(\vec{r})$
describes the generally position dependent transparency of the
tunneling contact at lead $\alpha$. In the following we choose the representation where
${\vec r}=(x,r_\perp)$, with $x$ being directed along the tube axis,
while $r_\perp$ is the coordinate on the nanotube cross-section. With $L$ being the
circumference of the SWNT ring, the contacts are positioned in the region about
 $x=0$ and $x=L/2$, see Fig.
\ref{ring}. Finally, ${\hat H}_{ext}$ accounts for the energy
dependence of the system on the external voltage sources
controlling the chemical potential in the leads: ${\hat
H}_{ext}=-e\sum_\alpha V_\alpha \sum_{{\vec
q}\sigma}c^\dagger_{\alpha {\vec q} \sigma} c_{\alpha {\vec q}
\sigma} $.
\section{Low energy description of metallic SWNT rings}\label{Sec2:low energy}
In this section we  derive the low energy Hamiltonian
$\hat{H}_{\mathrm{ring}}(\phi)=\hat{H}_{\rm kin} + \hat{V}_{\rm e-e} + \hat{H}_{\rm gate} $ of the SWNT ring and
diagonalize it. The ring Hamiltonian includes a kinetic term, the electron-electron interactions
as well as the effects of a capacitively applied gate voltage $V_g$ and of a magnetic flux $\phi$.
We assume a metallic
SWNT. Hence, depending on the diameter of
the SWNT, "low energies" means an energy range of the order of 1
eV around the Fermi energy, where the dispersion relation for the
non-interacting electrons near the Fermi points is linear and only
the two lowest subbands touching at the Fermi points can be
considered. The effects of the magnetic field are included by
introducing twisted boundary conditions (TBC) \cite{Loss93}.
Curvature and Zeeman effects are neglected here. Indeed the Zeeman splitting  yields a contribution   inversely proportional to the the square of the ring radius \cite{Lin98} and is relevant only for very small rings. Finally, the
linear dispersion relation around the Fermi points allows
bosonization \cite{Delft98} of the interacting Hamiltonian and
its successive diagonalization when only the forward scattering
part of the Coulomb interaction is included. The latter
approximation is justified for SWNTs with medium-to-large
cross-sections ($\emptyset>1.5$nm) \cite{Mayrhofer08}.
%
%the dispersion relation $\epsilon(\vec k)$ is linear. We shall see
%that the magnetic field can be taken account of by introducing
%twisted boundary conditions (TBC). Including the interactions we
%can then write down the Hamiltonian in second quantisation. In
%one-dimensional systems it is common practice to bosonize  the
%Hamiltonian as well as the electron operators $ $ which we will
%need for the calculation of the Green's functions. In order to
%study the time development of the bosonic operators we transform
%$\hat{H}_{\mathrm{ring}}$ into a diagonal form. This enables us to
%calculate the correlation functions of the electron operators,
%which are the main contribution to the Green's functions and thus
%to the conductance.
\subsection{Twisted boundary conditions }\label{quantization}
The band structure of non-interacting electrons in a SWNT ring is
conveniently derived from the band structure of the $p_z$
electrons in a graphene lattice. Since each unit cell of the
graphene lattice contains two carbon atoms, there are a valence
and a conduction band touching at the corner points of the first
Brillouin zone. Only two of these Fermi points, $\pm {\vec
K}_0=\pm\frac{4\pi}{3\sqrt{3}a_0}\hat{e}_x$,  are independent.
Since SWNTs are graphene sheets rolled up into a cylinder, we can
obtain the
%
 %Let us at first specify the quantization of the wave
%vector $\vec k$, starting from the graphene band structure.
%
electronic properties of a SWNT by imposing quantization of the
wave vector $\vec{k} = k_\bot \hat{e}_\bot + k_{||} \hat{ e}_{||}$
around the tube waist, i.e., denoting the SWNT circumference with
$L_\bot$, one finds $k_\bot = \frac{2\pi}{L_\bot}m,\ \ m=0,\pm
1,\pm 2\ldots\ ,$ which leads to the formation of several subbands
labelled by $m$. We consider here only SWNTs of the armchair type,
$\hat{e}_x=\hat{e}_{||}$, which are all metallic. In this case,
only the two sub-bands touching at the Fermi points $\pm
K_0\vec{e}_x$ are relevant.

\begin{figure}
    \centering
        \includegraphics{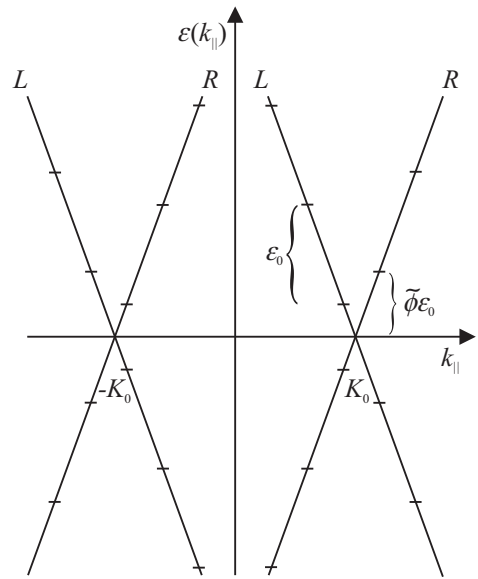}
    \caption{Low energy spectrum of a non-interacting SWNT ring threaded by a magnetic flux $\phi$.
    The underlying graphene structure is reflected in the two Fermi points at $\pm K_0$.
    Notice the misalignment $\tilde\phi\varepsilon_0$, with $\tilde\phi=\frac{\phi}{\phi_0} + \delta $, of the energy levels between the left ($L$) and right ($R$) branches due to
      the intrinsic mismatch $\delta$ and the applied flux  $\phi$. The flux quantum is denoted $\phi_0$.}
\label{disprel}
\end{figure}

Figure \ref{disprel} shows the linear valence and conduction bands
of an armchair SWNT. Corresponding to the direction of motion of
the electrons, there are two branches $b=R,L$ with positive and
negative slope, respectively.
The linear dispersion relation reads
\begin{eqnarray}
\varepsilon(b, \kappa) = \hbar v_F \mbox{sgn}(b) \kappa, \label{lin
disp}
\end{eqnarray}
where $\kappa$ measures the distance between $k_{||}$ and the
Fermi points $F=\pm K_0$, i.e., $k_{||}=\kappa+F$, and $v_F$ is
the Fermi velocity, $v_F = 8.1\times 10^5 $m/s.
Moreover, we use
the convention $b= r/l = \pm 1$.
The corresponding
 Bloch waves are of the form
\begin{eqnarray}
\varphi_{F b \kappa}(\vec{r}) &=& e^{i\kappa x} \varphi_{F
b}(\vec{r}) \label{bloch}
\\
&=& e^{i\kappa x} \frac{1}{\sqrt{N_L}}\sum_{\vec{R},p}f_{pFb}
e^{iFR_x} \Pi(\vec{r}-\vec{R}-\vec{\tau_p}). \nonumber
\end{eqnarray}
Here, $N_L$ is the number of lattice points in the nanotube lattice.
The index $p=1,2$ refers to the two graphene
sublattices and $\vec{\tau}_p$ points from a lattice point
$\vec{R}$ to a carbon atom of sublattice $p$. For an armchair
SWNT, the parameters $f_{pFb}$ are found to be $f_{1Fb} =
1/\sqrt{2},\ f_{2F b} = -1/\sqrt{2}\mbox{ sgn}(F b)$.  Finally, we have the $p_z$
electron orbital $\Pi$.

As we consider a ring, we can impose in the absence of a magnetic
field periodic boundary conditions along the direction parallel to
the tube axis, yielding   $k_{||}=\frac{2\pi}{L}m_k$, $m_k \in
\mathbb{Z}$. As the Fermi points $\pm K_0$ are not necessarily an
integer multiple of this step size, we introduce a mismatch
parameter $\delta$, $\delta \in [-0.5, 0.5)$, so that we can write
$K_0 = \frac{2\pi}{L}(n-\delta),\ \ n\in \mathbb{Z}.$ We still
have to take account of the magnetic field.  After travelling one
time around the ring, an electron picks up a phase
$2\pi\frac{\phi}{\phi_0}$, where $\phi_0 = \frac{h}{e}$ is the
flux quantum. This fact can be taken into account by introducing
twisted boundary conditions (TBC), $\Psi(\vec{r}+L\hat e_x)=e^{i2\pi\frac{\phi}{\phi_0}}\Psi(\vec{r})$ , instead of periodic ones
\cite{Loss93}. Under TBC the allowed values of the momentum parallel
to the tube axis $\kappa$ are
\begin{eqnarray}
\kappa &=& \frac{2\pi}{L}\left( n_\kappa + \frac{\phi}{\phi_0} + \delta \right),\ n_\kappa \in \mathbb Z \nonumber
\\
             &\equiv& \frac{2\pi}{L}\left( n_\kappa + \tilde{\phi} \right),\ m_\kappa \in \mathbb Z. \label{kappaquant}
\end{eqnarray}
Thus the electron operator $\Psi$  including the electron spin
$\sigma = \uparrow, \downarrow$ reads
\begin{eqnarray}
    \Psi(\vec{r}) &=& \sum_\sigma \Psi_\sigma(\vec r)= \sum_{F b \sigma\kappa}\varphi_{F b\kappa}^{}(\vec{r})
    c_{F b\sigma\kappa} \nonumber \\
                                &=& \sum_{F b\sigma\kappa}e^{i\kappa x}\varphi_{Fb}(\vec{r})c_{Fb\sigma\kappa}\nonumber
\equiv \sqrt{L}\sum_{Fb\sigma}\varphi_{Fb}(\vec{r})
\psi_{Fb\sigma}(x), \nonumber
                                                \\ \label{el op}
\end{eqnarray}
where $c_{Fb\sigma\kappa}$ annihilates an electron in state
$\left|Fb\sigma\kappa \right\rangle$ and $\psi_{Fb\sigma}(x)$ is a
slowly varying 1D operator. Later on the bosonisability of
$\psi_{Fb\sigma}(x)$ will be of importance to evaluate the
interference properties of the interacting SWNT ring.
\subsection{Kinetic energy and gate Hamiltonian}
The kinetic energy Hamiltonian in second quantization can in
principle be read off from the dispersion relation for the low
energy states (\ref{lin disp}):
\begin{eqnarray}
\hat{H}_{\rm kin} &=& \hbar v_F \sum_{F b \sigma
\kappa}\mbox{sgn}(b) \kappa c^\dagger_{F b\sigma\kappa}
c^{\phd}_{F b\sigma\kappa}.
\end{eqnarray}
Inserting the $\kappa$-quantization (\ref{kappaquant}),
introducing the number counting operator  $\mathcal N_{Fb\sigma} =
\sum_{\kappa} c^\dagger_{Fb\sigma\kappa}
c^{\phd}_{Fb\sigma\kappa}$ and using the abbrevation
$\varepsilon_0 \equiv \hbar v_F \frac{2\pi}{L}$ we get
\begin{eqnarray}
\hat{H}_{\rm kin} &=& \varepsilon_0 \sum_{Fb\sigma}\mbox{sgn}(b)
\left( \sum_\kappa m_\kappa c^\dagger_{Fb\sigma\kappa}
c^{\phd}_{Fb\sigma\kappa} + \tilde{\phi} \mathcal
N^{\phd}_{Fb\sigma}\right).  \nonumber
\\\label{kin energy}
\end{eqnarray}
Gate voltage effects are included in the Hamiltonian
${\hat H}_{\rm gate}=-\mu_{\rm g}{\cal N}=-e \alpha V_{\rm g}$,
where ${\cal N}=\sum_{Fb\sigma}\mathcal
N^{\phd}_{Fb\sigma}$ is the total electron operator of the SWNT,
$\mu_{\rm g}$ is the SWNT chemical potential and $\alpha$
a conversion factor.
\subsection{Interaction Hamiltonian} \label{Int Ham}
 A general form of the Hamiltonian ${\hat V}_{\rm e-e}$ describing
interactions between the electrons is
\begin{eqnarray}
&&{\hat V}_{\rm e-e} = \frac{1}{2}\sum_{\sigma \sigma'}
\\
&\times&  \int d^3r \int d^3r'
\Psi^\dagger_{\sigma}(\vec{r})\Psi^\dagger_{\sigma'}(\vec{r}\,')
U(\vec{r}-\vec{r}\,')
\Psi^{\phd}_{\sigma'}(\vec{r}\,')\Psi^{\phd}_{\sigma}(\vec{r}),
\nonumber
\end{eqnarray}
where $U(\vec{r}-\vec{r}\,')$ is the possibly screened Coulomb
potential. By expressing the 3D electron operator in terms of the
1D one, cf. Eq. (\ref{el op}), and integrating over the components
of $\vec{r}$ and $\vec{r}\,'$ perpendicular to the tube axis one
obtains an effective one-dimensional interaction. As discussed in
\cite{Egger97}, this yields  in general  interlattice and
intralattice interactions. In \cite{Mayrhofer08} it is shown that
when the tube diameter is large enough, the distinction between
intra and inter-lattice interactions is no longer relevant, and
only forward scattering processes, where the number of electrons
in each branch remains constant, are relevant.

 Introducing the 1D electron density operator
$\rho^{\phd}_{Fb\sigma}(x) = \psi^\dagger_{Fb\sigma}(x)
\psi^{\phd}_{Fb\sigma}(x)$ and retaining only forward scattering
processes we obtain for the interaction Hamiltonian
\begin{eqnarray}
{\hat V}_{\rm e-e}&\approx &  \frac{1}{2} \sum_{FF'}\sum_{bb'} \sum_{\sigma
\sigma'} \int_0 ^L dx \int_0 ^L dx'  \nonumber
\\
&\times & \rho_{Fb\sigma}(x) V_0(x, x') \rho_{F'b'\sigma'}(x'),
\label{V dens}
\end{eqnarray}
with the effective 1D potential $V_0(x, x')$ defined as
\begin{eqnarray}
\lefteqn{V_0(x, x') =  \frac{L^2}{N_L^2}\sum_{\vec{R},\vec{R'}}\int
d^2r_\bot \int d^2r'_\bot \times}\nonumber
\\
&\times &  |\Pi(\vec{r} -\vec{R})|^2 U(\vec{r}-\vec{r}\,'
)|\Pi(\vec{r}\,' -\vec{R'})|^2.
\end{eqnarray}
\subsection{Bosonization}
The problem of a one-dimensional system of interacting electrons is often conveniently
formulated in terms of bosonic operators. A didactic overview
can be found in \cite{Delft98}. To this extent we assume a linear
 dispersion relation over
all values of the momentum $\kappa$. This is justified by the
long-range character of the interaction, leading to a momentum
cut-off still lying in the low-energy range around the Fermi
energy. The states with negative energy are assumed to be filled
in the ground state, the Fermi sea, of the system. As in
\cite{Delft98}, we now introduce new bosonic operators
\begin{eqnarray}
b_{F\sigma q} &:=& \frac{1}{\sqrt{|n_q|}} \left\{ \begin{array}{c}\sum_\kappa c^\dagger_{Fr\sigma\kappa}c^{\phd}_{Fr\sigma\kappa+q},\ \ \ q>0\phantom{,} \\ \sum_\kappa c^\dagger_{Fl\sigma\kappa}c^{\phd}_{Fl\sigma\kappa+q},\ \ \ q<0, \end{array} \right. \label{b}
\end{eqnarray}
with $q= \frac{2\pi}{L}n_q, n_q \in \mathbb Z$, and obeying the canonical bosonic commutation relations
\begin{eqnarray}
[b^{\phd}_{F_1\sigma_1 q_1}, b^\dagger_{F_2\sigma_2 q_2}] = \delta_{F_1F_2}\delta_{\sigma_1\sigma_2}\delta_{q_1q_2},\ [b^{\phd}_{F_1\sigma_1 q_1}, b^{\phd}_{F_2\sigma_2 q_2}] = 0. \nonumber
\\
\end{eqnarray}
%
%At first we give the result for the bosonized electron operator. A derivation is shown e.g. in \cite{delft}. The electron operator is written as a product of a fermionic operator, the \emph{Klein factor}, and a bosonic field operator. The former lowers the number of electrons with given quantum numbers $Fr\sigma$ by one.
%\begin{eqnarray}
 %   \psi_{Fr\sigma}(x) &=& \eta_{Fr\sigma}e^{i\frac{2\pi}{L}\left(\mathcal N_{Fr\sigma} \mbox{\scriptsize sgn}(r) +\tilde{\phi}\right) x}e^{i\phi^\dagger_{Fr\sigma}(x)}e^{i\phi^{\phd}_{Fr\sigma}(x)},\label{psi boson}\nonumber
%\\
%\\ \label{phi}
 %   \phi^\dagger_{Fr\sigma}(x) &=& i\sum_{q>0}\frac{1}{\sqrt{|n_q|}} e^{-i\mbox{\scriptsize sgn}(r) %qx}b^\dagger_{F\sigma \mbox{\scriptsize sgn}(r)q} e^{-\alpha q/2}.
%\end{eqnarray}
%The infinitesimally small number $\alpha>0$ in (\ref{phi}) serves to avoid divergences due to the extension of the %linearised dispersion relation.
%\\
In terms of the bosonic operators (\ref{b})  the kinetic energy Hamiltonian (\ref{kin energy}) assumes the form
\begin{eqnarray}
\hat{H}_{ \mathrm{kin}}    &:=&\varepsilon_{0} \sum_{F \sigma}\sum_{q\neq 0} |n_q| b^\dagger_{F\sigma q} b^{\phd}_{F\sigma q}  + \nonumber
\\
&+& \varepsilon_{0} \sum_{Fb\sigma} \left[ \frac{\mathcal N^2_{Fb\sigma}}{2}+  \tilde{\phi}\, \mbox{sgn}(b) \mathcal N_{Fb\sigma}\right].  \label{H0boson}
\end{eqnarray}
For the bosonization of the interaction part (\ref{V dens}), we decompose the density operator $\rho_{Fb\sigma}$ into its Fourier components, which are proportional to  the bosonic operators (\ref{b}):
\begin{eqnarray}
&& \rho^{\phd}_{Fb\sigma}(x) = \psi^\dagger_{Fb \sigma}(x) \psi^{\phd}_{Fb \sigma}(x) = \mathcal{N}_{Fb\sigma} \label{rho}
\\
&&+ \sum_{q > 0} \sqrt{|n_q|} \left( e^{i \mbox{\scriptsize sgn}(b) qx} b^{\phd}_{F\sigma \mbox{\scriptsize sgn}(b) q} +  e^{-i\mbox{\scriptsize sgn}(b) qx}  b^\dagger_{F \sigma \mbox{\scriptsize sgn}(b) q} \right). \nonumber
\end{eqnarray}
By inserting (\ref{rho}) into (\ref{V dens}) we get
\begin{eqnarray}
\hat{V} &=& \frac{1}{2} W_{00} \mathcal N^2 + \sum_{F\sigma} \sum_{F'\sigma'}\sum_{q \neq 0}\frac{1}{2}  W_{q -q} |n_q| \times \nonumber\\
&\times &\left(b^{\phd}_{F\sigma q} + b^\dagger_{F\sigma -q} \right) \left(b^{\phd}_{F'\sigma' -q} + b^\dagger_{F'\sigma' q} \right), \nonumber
\end{eqnarray}
with the total number of electrons $\mathcal N = \sum_{Fb\sigma} \mathcal N_{Fb\sigma}$ in the ring. The effective interaction $V_0(x,x')$ is absorbed into
\begin{eqnarray}
W_{qq'}        &\equiv & \int_0 ^L dx \int_0 ^L dx'  e^{i(qx +q'x')} V_0(x, x').
\end{eqnarray}
This is nonzero only for $q'=-q$ and yields
\begin{eqnarray}
W_{q-q} =L \cdot \int_{-L/2}^{L/2} dy \cos(qy)V_0(|y|).
\end{eqnarray}
The parameter $W_{00}$ can be identified with the charging energy $E_c$ responsible for Coulomb blockade.
To proceed, it is convenient to introduce linear combinations of the bosonic operators (\ref{b})
associated to spin, charge and orbital degrees of freedom:
\begin{eqnarray}
b_{\oplus cq}   &:=& \frac{1}{2}(b_{K_0 \uparrow q}+ b_{K_0 \downarrow q} + b_{-K_0 \uparrow q}+ b_{-K_0 \downarrow q}),
\nonumber\\
 b_{\ominus cq} &:=& \frac{1}{2}(b_{K_0 \uparrow q}+ b_{K_0 \downarrow q} - b_{-K_0 \uparrow q}- b_{-K_0 \downarrow q}),
\nonumber\\
b_{\oplus sq} &:=& \frac{1}{2}(b_{K_0 \uparrow q}- b_{K_0 \downarrow q} + b_{-K_0 \uparrow q}- b_{-K_0 \downarrow q}),
\nonumber\\
b_{\ominus sq} &:=& \frac{1}{2}(b_{K_0 \uparrow q}- b_{K_0 \downarrow q} - b_{-K_0 \uparrow q}+ b_{-K_0 \downarrow q}), \nonumber
 %\label{b-sq}
\end{eqnarray}
or shortly: $b_{\tilde F j q}$, $\tilde F = \oplus / \ominus$, $j =c,s$. The advantage of this transformation is that
 the term quadratic in the bosonic operators in $\hat V_{\rm e-e}$ contains now only  $b_{\oplus c q}$-type operators.
 The ring Hamiltonian  finally reads
\begin{eqnarray}
&& \hat{H}_{\mathrm{ring}}  = \varepsilon_{0} \sum_{\tilde{F}j} \sum_{q\neq 0} |n_q| b^\dagger_{\tilde{F} jq} b^{\phd}_{\tilde{F} jq}  -\mu_{\rm g}{\cal N}  \nonumber
\\
&+& \varepsilon_0 \sum_{Fb\sigma} \left( \frac{\mathcal N^2_{Fb\sigma}}{2}+  \tilde{\phi}\, \mbox{sgn}(b) \mathcal N_{Fb\sigma}\right) + \frac{1}{2} W_{00} \mathcal{N}^2  \nonumber
\\
&+& \frac{1}{2} \sum_{q\neq0} 4 |n_q| W_{q -q}  \left(b^{\phd}_{\oplus cq} +  b^\dagger_{\oplus c-q}\right) \left( b^{\phd}_{\oplus c-q} + b^\dagger_{\oplus cq}\right).  \nonumber
\\
\label{V boson sc}
\end{eqnarray}
\subsubsection{Diagonalization}
The Hamiltonian (\ref{V boson sc}) can be diagonalized by a Bogoliubov transformation. More details can be found in the Appendix. As a result we get
\begin{eqnarray}
\lefteqn{\hat{H}_\mathrm{ring} = \sum_{\tilde{F}j}\sum_{q \neq 0}  \varepsilon_{\tilde{F}jq} a_{\tilde{F}jq}^\dagger a^{\phd}_{\tilde{F}jq} + \varepsilon_0 \sum_{F b \sigma} \frac{\mathcal{N}^{ 2}_{Fb\sigma}}{2}} \nonumber
\\
&+& \varepsilon_0 \tilde{\phi} \sum_{Fb\sigma} \mbox{sgn}(b) \mathcal{N}_{Fb\sigma}
+ \frac{1}{2} W_{00} \mathcal{N}^2
-\mu_{\rm g}{\cal N}.  \; \label{Ham}
\end{eqnarray}
The definition of the energies $\varepsilon_{\tilde F jq}$ as well as the relation between the new ($a_{\tilde{F}jq}$) and the old ($b_{\tilde{F}jq}$) operators can be found in the Appendix. The degrees of freedom which are affected by the interaction are those related to the $a_{\oplus c q}$-operators. Therefore, these are called \emph{charged} degrees of freedom, whereas the indices $(\ominus cq)$, $(\oplus sq)$ and $(\ominus sq)$ denote the \emph{neutral} modes. In total, there are one charged and three neutral modes.
%remember that $q$ can be positive or negative).
\\
In order to investigate the combined effects of gate voltage and magnetic flux, it is convenient to introduce
appropriate linear combinations of the number counting operator $\mathcal N_{Fr\sigma}$ :
\begin{eqnarray}
%\begin{array}{rrllllll}
\mathcal N& \equiv &\mathcal N_{\oplus c}  :=  \sum\limits_{F b\sigma}  \mathcal N_{F b\sigma},
\nonumber\\
\mathcal J &\equiv& \mathcal J_{\oplus c} := \sum\limits_{F b\sigma} \mbox{sgn}(b) \mathcal N_{F b\sigma},
\end{eqnarray}
 being the total particle and total current operators and
\begin{eqnarray}
%\begin{array}{llllll}
\mathcal N_{\oplus s} &:=&  \sum\limits_{F b\sigma} \mbox{sgn}(\sigma) \mathcal N_{F b\sigma},
\qquad \mathcal J_{\oplus s} :=  \sum\limits_{F b\sigma} \mbox{sgn}(b\sigma)\mathcal N_{F b\sigma}, \nonumber \\
\mathcal N_{\ominus c} &:=& \sum\limits_{F b\sigma} \mbox{sgn}(F)\mathcal N_{F b\sigma},\qquad
\mathcal J_{\ominus c} := \sum\limits_{F b\sigma} \mbox{sgn}(Fb)\mathcal N_{Fb\sigma},
\nonumber\\
\mathcal N_{\ominus s} &:=&\sum\limits_{F b\sigma} \mbox{sgn}(F\sigma)\mathcal N_{F b\sigma},
\quad\mathcal J_{\ominus s} := \sum\limits_{Fb\sigma} \mbox{sgn}(Fb\sigma)\mathcal N_{Fb\sigma}.\nonumber
%\end{array}
\end{eqnarray}
Expressing $\hat{H}_\mathrm{ring}$ in terms of these newly defined operators yields
\begin{eqnarray}
\hat{H}_{\mathrm{ring }} &:=&  \hat{H}_\mathrm{ring,b}+\hat{H}_\mathrm{ring,f}= \sum_{\tilde{F}j}\sum_{q \neq 0}  \varepsilon^{\phd}_{\tilde{F}jq} a_{\tilde{F}jq}^\dagger a^{\phd}_{\tilde{F}jq} \nonumber \\
&+&
\frac{\varepsilon_0}{16} [\tilde{W} ({\mathcal N}-k_{\mu_g})^2   + ({\mathcal J}-k_\phi)^2]
\nonumber \\&+&
\frac{\varepsilon_0}{16}\sum_{\tilde F j\neq \oplus c}(\mathcal N^2_{\tilde F j} + \mathcal J^2_{\tilde F j})
+ {\rm const},\label{H diag}
\end{eqnarray}
where $\tilde{W} = 1+ \frac{8W_{00}}{\varepsilon_0}$,
$k_{\mu_g} = \frac{\mu_g}{W_{00} + \varepsilon_0 /8}$ and $k_\phi = -8 \tilde \phi$.
%
%\begin{eqnarray}
%&& \hat{H}_\mathrm{ring} = \sum_{\tilde{F}j}\sum_{q \neq 0}  \varepsilon^{\phd}_{\tilde{F}jq} a_{\tilde{F}jq}^\dagger a^{\phd}_{\tilde{F}jq}  + \nonumber
%\\
%&&+ \frac{1}{2} W_{00} \mathcal{N}^2+ \frac{\varepsilon_0}{16} \sum_{\tilde F j}(\mathcal N^2_{\tilde F j} + \mathcal %J^2_{\tilde F j})- \mu_g \mathcal N + \varepsilon_0 \tilde{\phi}  \mathcal J .\nonumber
%\\ \label{H diag}
%\end{eqnarray}
The first line corresponds to the bosonic part of the Hamiltonian, $\hat{H}^{\phd}_{\mathrm{ring, b}}$ and the second and third lines to the fermionic part $\hat{H}^{\phd}_{\mathrm{ring, f}}$.
An eigenbasis of $\hat H_{\rm ring}$ is thus formed by the states
\begin{eqnarray}
\left \{ \prod_{ \tilde{F}j,q \neq 0,} \frac{(a^\dagger_{\tilde{F} j q})^{m_{\tilde{F} jq}}}{\sqrt{m_{\tilde{F} jq}!}} \left|\vec{N}\right\rangle \right \}   =: \left \{\left|\vec{N}, \vec{m}\right\rangle \right \},
\label{eigenstates}
\end{eqnarray}
where $\left|\vec{N}, \vec{0}\right\rangle $ has no bosonic excitations.
  In general, any number of electrons
 can be distributed in many different ways on the branches $(F b \sigma)$, which is described by the set of vectors
\begin{eqnarray}
\vec{N}&=&(N_{K_0 r\uparrow}, N_{K_0 r\downarrow}, N_{K_0 l\uparrow}, N_{K_0 l\downarrow}, \nonumber
\\ && N_{-K_0 r\uparrow}, N_{-K_0 r\downarrow}, N_{-K_0 l\uparrow}, N_{-K_0 l\downarrow}).
\end{eqnarray}
For each value of $\vec{N}$ one has to include  the  states which contain
the bosonic excitations of the interacting electrons, where
the the parameter $m_{\tilde{F} jq}$ counts the number of  excitations in the channel $\tilde F j$ with momentum $q$.
\section{Linear transport}\label{Sec3:linear transport}
We have now all the ingredients to evaluate the linear transport characteristics of the
interacting SWNT ring by use of the Kubo formula.
\subsection{Conductance formula}\label{cond form}
The current operator at lead $L$ is defined as the rate of change
of particles at the  contact $L$, i.e., $\hat I_L=-e {\dot{\hat
N}}_L=-\frac{ie}{\hbar}[\hat H,\hat N_L]=-\frac{ie}{\hbar}[\hat
H_{TL},\hat N_L]$ where $-e$ is the electron charge. It yields
\begin{equation}
\hat I_L = -\frac{ie}{\hbar} \sum_\sigma \int d^3 r (T_L(\vec r)
\Psi_\sigma ^\dagger (\vec r) \Phi_{\sigma L}(\vec r) - h.c.).
\end{equation}
Analogously  the current operator at the right lead is
\begin{equation}
 \hat I_R  = \frac{ie}{\hbar} \sum_\sigma \int d^3 r
(T_R(\vec r) \Psi_\sigma ^\dagger (\vec r) \Phi_{\sigma R}(\vec r)
- h.c.).\;
\end{equation}
To proceed, we use relation (\ref{el op}), relating the 3D operator $\Psi(\vec{r})$ to the slowly varying 1D one
$\psi_{Fb\sigma}(x)$.
We assume that the latter does not change significantly in the tunneling region.  This yields for the current operators
\begin{eqnarray}
\lefteqn{\hat I_L     = -\frac{ie}{\hbar} \sum_{Fb\sigma}\sum_{\vec q}\times}
 \nonumber \\
&\hspace{-1cm} &\hspace{-1cm}
 \Biggl( \underbrace{\sqrt{L}\int d^3 r T_L(\vec r) \varphi^*_{Fb}(\vec r) \phi_{\vec q, L} (\vec r)}_{=: T_{LFb\vec q }}\psi_{Fb\sigma}^\dagger (x_L) c_{\vec q \sigma L} - h.c.\Biggr) \nonumber
\\
&=& -\frac{ie}{\hbar} \sum_{Fb\sigma}\sum_{\vec q} \Biggl(T_{L Fb \vec q}\psi_{Fb\sigma}^\dagger (x_L) c_{\vec q \sigma L} - h.c.\Biggr),
\\
\hat I_R     &=& \frac{ie}{\hbar} \sum_{Fb\sigma}\sum_{\vec q} \Biggl(T_{R Fb\vec q}\psi_{Fb\sigma}^\dagger (x_R) c_{\vec q \sigma L} - h.c.\Biggr),
\end{eqnarray}
where $x_L$ and $x_R$ are in the middle of the respective tunneling regions.
The above expressions for the current operators can in turn be used to
%Without considering the interactions among the electrons in the
%ring, an expression for the conductance $G$ in a closed form can
%be found using the Landauer formula \cite{buttiker}.
  evaluate the DC-conductance of the interacting SWNT ring in terms
of the Kubo formula
\begin{eqnarray}
G = \lim_{\Omega \rightarrow 0} \mbox{Re} \frac{1}{\hbar
\Omega}\int_{- \infty}^t dt' e^{i \Omega (t'-t)} \langle [
\hat{I}_L( t),\ \hat{I}_R (t')] \rangle_{\rm eq}, \label{conductance def}
\end{eqnarray}
with $\hat{I}_L (t)$ and $\hat{I}_R (t)$ being current operators in
the interaction representation, where the interaction is
represented by the Hamiltonian ${\hat H}_{ext}$. Hence, $\langle
\quad\rangle_{\rm eq}$ describes the average with respect to the
equilibrium density operator ${\hat \rho}_{\rm eq} := Z^{-1}e^{-\beta({{\hat
H}_{\rm ring}+ {\hat H}_T + {\hat H}_L + {\hat H}_R})}$, with $\beta$ the inverse temperature and
$Z$ the partition function.
It is convenient to introduce  the linear susceptibility $\chi(t)=-\frac{i}{\hbar}\theta(t)\langle [
\hat{I}_L( t),\ \hat{I}_R (0)] \rangle_{\rm eq}$ in terms of which (\ref{conductance def})  assumes the
compact form
\begin{eqnarray}
G = \frac{1}{\hbar}\lim_{\Omega \rightarrow 0}i\frac{\tilde\chi(\Omega)}{\Omega}.
\label{conductance-chi}
\end{eqnarray}
Here
$\tilde\chi(\Omega)$ is the Fourier transform
 of the response function $\chi(t)$.  In \cite{Kinaret98} the linear susceptibility at imaginary times
\begin{equation}
\chi(\tau_1 - \tau_2)=-\langle T_\tau [\hat I_L(\tau_1),\hat I_R(\tau_2)]\rangle_{\rm eq},
\end{equation}
where $T_\tau$ indicates time-ordering,
has been evaluated  for a spinless Luttinger liquid.
 Generalizing \cite{Kinaret98} to the multichannel situation represented by a 3D toroidal SWNT, an expression for the linear susceptibility
 can be obtained to lowest non vanishing order in the tunneling matrix elements
$T_{\alpha F b \vec{q}}$. A summary of the calculation, where we made use of the explicit form of the nanotube Bloch wave functions, can be found in the subsection below. It delivers for the conductance the remarkably simple result
\begin{eqnarray}
G &\approx& \frac{e^2}{h}   \frac{|\Phi_L |^2|\Phi_R|^2}{\hbar^2}   \label{cond final}\\
&\times & \sum_{Fb\sigma}
 \int_{-\infty}^{+\infty} d\omega \Biggl(-\frac{\partial n_{\rm F}(\omega)}{\partial \omega} \Biggr)|G^{\rm ret}_{Fb\sigma}(\omega, x_L - x_R)|^2, \nonumber
\end{eqnarray}
where $n_{\rm F}(\omega)$ is the Fermi function and
$G^{\rm ret}_{Fb\sigma}(\omega, x_L - x_R)$,
the retarded Green's function for the interacting electrons on the SWNT ring, is entirely determined by the slowly varying  1D part of the electron operator (\ref{el op}) as it is the Fourier transform of
\begin{eqnarray}
 G^{\rm ret}_{Fb\sigma} (t, {x} - {x}\,') &=& -\frac{i}{\hbar}  \theta (t) \left\langle \{ \psi_{Fb\sigma}({x}, t) , \psi^\dagger_{Fb\sigma}({x}\,', 0) \} \right\rangle . \nonumber \\
\label{c6 green}
\end{eqnarray}
As the number $\cal N$ of electrons in the dot can vary,
 $\langle \ \rangle$ indicates the thermal average with respect to the grandcanonical equilibrium density matrix $\rho_{{\rm ring},{\cal N}}=Z^{-1}_{\rm ring}\exp^{-\beta \hat H_{\rm ring}}$.
 In contrast, knowledge of the 3D character of the SWNT Bloch wave functions is encapsulated in the tunneling functions
 $\Phi_R$ and $\Phi_L$, see Eq. (\ref{transmission}) below.
    We notice that Eq. (\ref{cond final}) is not trivial, as it predicts the \emph{absence of
    interference} between Green's functions with different indices, in contrast to the case of a strictly one-dimensional ring considered in \cite{Kinaret98} and the formula given in \cite{Rollbühler99}. This result has its origin in the
    very strongly localized character of the $p_z$ orbitals, and on the fact that we assumed extended contacts
     coupling equally to both sublattices of the underlying graphene structure.
     For temperatures further than $k_BT$ from a resonance (\ref{cond final}) further simplifies to
     \begin{equation}
G \approx \frac{e^2}{h}   \frac{|\Phi_L |^2|\Phi_R|^2}{\hbar^2}
 \sum_{Fb\sigma}
|G^{\rm ret}_{Fb\sigma}(0, x_L - x_R)|^2. \label{cond zero-temperature}
\end{equation}
\subsection{Proof of the conductance formula (\ref{cond final})}
In this subsection,
  which the hurried reader can skip,  the linear susceptibility $\tilde\chi(\Omega)$ entering the conductance formula
  (\ref{conductance-chi}) is obtained by first evaluating the Fourier transform $\tilde\chi(i\Omega_n)$ of the imaginary
   time response function $\chi(\tau)$ and successive analytic continuation: $\tilde\chi(\Omega)=\lim_{i \Omega_n\to \Omega} \tilde\chi(i\Omega_n)$. Specifically, we use the generating function method to find an expression for  the imaginary time susceptibility at lowest order in the tunneling couplings \cite{Kinaret98}. It reads
\begin{widetext}
\begin{eqnarray}
&&  \chi (\tau_1 - \tau_2) =  - \frac{e^2}{\hbar^4} \int_0^{\hbar\beta} d\tau \int_0^{\hbar\beta} d\tau' \sum_{F_1 b_1} \sum_{F_2 b_2} \sum_{F_3 b_3} \sum_{F_4 b_4} \sum_{\sigma \sigma'} \sum_{\vec q \vec q'}  T_{L F_1 b_1 \vec q} T^*_{L F_2 b_2 \vec q} T_{R F_3 b_3 \vec q'} T^*_{R F_4 b_4 \vec q'}  \times
\nonumber   \\
&& \times \Bigl \{
 \langle T_\tau\left[\psi^\dagger_{F_1 b_1 \sigma}(\tau, x_L) \psi_{F_2 b_2 \sigma}(\tau_1, x_L) \psi^\dagger_{F_3 b_3 \sigma'}(\tau', x_R) \psi_{F_4 b_4 \sigma'}(\tau_2, x_R) \right] \rangle G_{L \sigma}(\vec q, \tau-\tau_1) G_{R \sigma}(\vec q',\tau'-\tau_2)+ \nonumber
\\&&  \phantom{\ }
- \langle T_\tau\left[\psi^\dagger_{F_1 b_1 \sigma}(\tau, x_L) \psi_{F_2 b_2 \sigma}(\tau_1, x_L) \psi^\dagger_{F_3 b_3 \sigma'}(\tau_2, x_R) \psi_{F_4 b_4 \sigma'}(\tau', x_R) \right]\rangle G_{L \sigma}(\vec q,\tau-\tau_1) G_{R \sigma}(\vec q',\tau_2-\tau')+ \nonumber
\\&& \phantom{\ }
-  \langle T_\tau\left[\psi^\dagger_{F_1 b_1 \sigma}(\tau_1, x_L) \psi_{F_2 b_2 \sigma}(\tau, x_L) \psi^\dagger_{F_3 b_3 \sigma'}(\tau', x_R) \psi_{F_4 b_4 \sigma'}(\tau_2, x_R) \right]\rangle G_{L \sigma}(\vec q,\tau_1-\tau) G_{R \sigma}(\vec q',\tau'-\tau_2)+ \nonumber
\\&& \phantom{\ }
+  \left. \langle T_\tau\left[\psi^\dagger_{F_1 b_1 \sigma}(\tau_1, x_L) \psi_{F_2 b_2 \sigma}(\tau, x_L) \psi^\dagger_{F_3 b_3 \sigma'}(\tau_2, x_R) \psi_{F_4 b_4 \sigma'}(\tau', x_R) \right]\rangle G_{L \sigma}(\vec q,\tau_1-\tau) G_{R \sigma}(\vec q',\tau_2-\tau') \right \},
\label{chi}
\end{eqnarray}
   \end{widetext}
   where $G_{\alpha \sigma}(\vec q, \tau-\tau') = -\langle T_\tau (c_{\vec q \sigma \alpha} (\tau) c^\dagger_{\vec q \sigma \alpha}(\tau'))\rangle_\alpha$ denotes the Green's function for the free electrons of lead $\alpha$.
   In (\ref{chi}) $\langle
\quad\rangle_\alpha$ describes the average with respect to the
equilibrium density operator $Z^{-1}_\alpha e^{-\beta{\hat H}_\alpha }$ of the lead $\alpha$.

Though the four particles correlator can in principle be  evaluated using the bosonisation approach, see e.g. \cite{Pletyukhov06}, following \cite{Kinaret98} we  factorize it as
\begin{widetext}
\begin{eqnarray}
\lefteqn{   \langle T_\tau\psi^\dagger_{F_1 b_1 \sigma}(\tau, x_L) \psi_{F_2 b_2 \sigma}(\tau_1, x_L) \psi^\dagger_{F_3 b_3 \sigma'}(\tau', x_R) \psi_{F_4 b_4 \sigma'}(\tau_2, x_R) \rangle }\nonumber
    \\
    &\approx& \delta_{F_1 F_2} \delta_{b_1 b_2} \delta_{F_3 F_4} \delta_{b_3 b_4}\langle T_\tau \psi_{F_2 b_2 \sigma}(\tau_1, x_L) \psi^\dagger_{F_1 b_1 \sigma}(\tau, x_L)\rangle \langle \psi_{F_4 b_4 \sigma'}(\tau_2, x_R)   \psi^\dagger_{F_3 b_3 \sigma'}(\tau', x_R) \rangle\nonumber
    \\
    &-& \delta_{F_2 F_3} \delta_{b_2 b_3} \delta_{F_4 F_1} \delta_{b_2 b_3}\delta_{\sigma \sigma '}\langle T_\tau \psi_{F_2 r_2 \sigma}(\tau_1, x_L) \psi^\dagger_{F_3 b_3 \sigma'}(\tau', x_R)\rangle \langle \psi_{F_4 b_4 \sigma'}(\tau_2, x_R) \psi^\dagger_{F_1 b_1 \sigma}(\tau, x_L) \rangle.
\end{eqnarray}
\end{widetext}
In this approximation multiple interference processes are neglected. The conductance is thus expressed in terms of the single particle Green's functions
\begin{equation}
G_{Fb\sigma}(\tau-\tau',x_\alpha -x_\beta):=\langle T_\tau \psi_{F b \sigma}(\tau, x_\alpha) \psi^\dagger_{F b \sigma}(\tau', x_\beta)\rangle.
\end{equation}
In frequency space this yields
\begin{eqnarray}
    \lefteqn{\tilde\chi(i\Omega_n) = -\frac{e^2}{\hbar^4} \frac{1}{\hbar \beta}\sum_{i\omega_n, i\omega'_n}  \delta_{\Omega_n, \omega_n - \omega'_n }}
  \nonumber \\
  &\times&\sum_{Fb}\sum_{F'b'}\sum_\sigma \sum_{\vec q \vec q'}  T_{L Fb \vec q} T^*_{L F'b' \vec q} T_{R F'b' \vec q'} T^*_{R Fb \vec q'}\nonumber \\
    &\times &  G_{F'b'\sigma}(i\omega_n, x_L - x_R) G_{Fb\sigma}(i\omega_n-i\Omega_n, x_R-x_L)   \nonumber \\
&\times& \left[G_{L \sigma}(\vec q, i\omega_n - i\Omega_n) - G_{L \sigma}(\vec q, i\omega_n) \right]
 \nonumber \\
 &\times&
 \left[G_{R \sigma}(\vec q', i\omega'_n) - G_{R \sigma}(\vec q', i\omega'_n + i\Omega_n) \right].
\end{eqnarray}
 As a further step  we perform the summation over the fermionic frequencies and  carry out the analytical continuation
 $i\Omega_n\to \Omega$ to find $\tilde\chi (\Omega)$.
 From (\ref{conductance-chi}) the conductance then follows as
\begin{eqnarray*}
\lefteqn{G
=\frac{1}{h}  \frac{e^2}{\hbar^4}  \sum_{Fb}\sum_{F'b'}\sum_\sigma \sum_{\vec q \vec q'}  T_{L Fb \vec q} T^*_{L F'b' \vec q} T_{R F'b' \vec q'} T^*_{R Fb \vec q'}}\\
&\times&    \int_{-\infty}^{\infty}d\omega G^{\rm ret}_{Fb\sigma}(\omega, x_L - x_R) G^{\rm adv}_{F'b'\sigma}(\omega, x_R - x_L)
\\ && \times \Biggl(-\frac{\partial n_F(\omega)}{\partial\omega}\Biggr)4\mbox{Im}G^{\rm ret}_{L \sigma}(\vec q, \omega ) \mbox{Im}G^{\rm ret}_{R \sigma}(\vec q', \omega),
\end{eqnarray*}
where the superscript "ret/adv" refer to retarded/advanced Green's functions, respectively.
Such expression can be simplified further by replacing the summation over the $\vec q$-values with an integral over energies: $\sum_{\vec q} = \int d\varepsilon \rho_L(\varepsilon) \sum_{\vec q_{|\varepsilon}}$,
where $\rho_L(\varepsilon)$ is the density of states of lead $L$. Analogous procedure holds for the right lead.
Recalling the explicit expression of the tunneling function $T_{\alpha F b \vec q}$ one then finds
\begin{eqnarray}
G &=& \frac{L^2}{h}  \frac{e^2}{\hbar^4}  \sum_{Fb}\sum_{F'b'}\sum_\sigma \int d\varepsilon \rho_L(\varepsilon) \sum_{\vec q_{|\varepsilon}} \int d\varepsilon' \rho_R(\varepsilon') \sum_{\vec q'_{|\varepsilon'}}\nonumber
\\
&\times &\int d^3 r_1 T_L(\vec r_1) \varphi^*_{Fb}(\vec r_1) \frac{1}{\sqrt{V_{L}}} e^{i \vec q_{|\varepsilon} \cdot \vec r_1}
                     \nonumber \\
                     &\times& \int d^3 r_2 T^*_L(\vec r_2) \varphi_{F'b'}(\vec r_2)\frac{1}{\sqrt{V_{L}}} e^{-i \vec q_{|\varepsilon}\cdot \vec r_2 } \label{conductance-int}
\\
&\times & \int d^3 r_3 T_R(\vec r_3) \varphi^*_{F'b'}(\vec r_3)\frac{1}{\sqrt{V_{R}}} e^{i \vec q_{|\varepsilon'}\cdot \vec r_3} \nonumber
\\ &\times &        \int d^3 r_4 T^*_R(\vec r_4) \varphi_{Fb}(\vec r_4) \frac{1}{\sqrt{V_{R}}}e^{-i \vec q_{|\varepsilon'}\cdot \vec r_4}\nonumber
    \\
&\times &   \int_{-\infty}^{\infty}d\omega G^{\rm ret}_{Fb\sigma}(\omega, x_L - x_R) G^{\rm adv}_{F'b'\sigma}(\omega, x_R - x_L)
\nonumber \\ && \times  \Biggl(- \frac{\partial n_F(\omega)}{\partial\omega}\Biggr) 4 \pi^2 \delta(\omega-\varepsilon/\hbar) \delta(\omega-\varepsilon'/\hbar),\nonumber
\end{eqnarray}
where we made use of the expression for the spectral density of a free electron gas $A_{\alpha\sigma}(\vec q, \omega) := -2 \mbox{Im}\left(G^{\rm ret}_{\alpha \sigma}(\vec q,\omega)\right) = 2\pi\delta(\omega-\varepsilon_{\alpha \vec q}/\hbar)$,
and we assumed that the leads wave functions are well approximated by plane waves.
A crucial simplification follows now from the observation that the $p_z$ orbitals, entering the products $\varphi^*_{Fb}(\vec r_1)\varphi_{Fb}(\vec r_4)$, see (\ref{bloch}), are strongly localized at the carbon lattice sites.
In contrast,  the other
position-dependent functions entering (\ref{conductance-int}) are slowly varying on the scale of the extension of the
localized $p_z$ orbitals. Hence, we can replace the latter with delta-functions centered at the position
$\vec{R}+\vec{\tau}_p$ of the carbon atoms.
The summation over the wave numbers $\vec{q}$ associated with an energy $\varepsilon$ yields
\begin{eqnarray}
\lefteqn{\sum_{\vec q_{|\varepsilon}} \frac{1}{V_L} e^{i\vec q_{|\varepsilon}(\vec R +\vec{\tau}_p- \vec{R''}-\vec{\tau}_{p''})} } \\& =
&\frac{4\pi \sin(q_{|\varepsilon}|\vec{R}+\vec{\tau}_p - \vec {R''}-\vec{\tau}_{p''}|)}{q_{|\varepsilon} |\vec{R}+\tau_p - \vec R''-\vec\tau_p{''}|} \approx 4\pi \delta_{\vec R,\vec R''}\delta_{p,p''},\nonumber
\end{eqnarray}
where the latter equation means that we assume the  lead wave vectors $\vec{q}_{|\varepsilon_F} $ at the Fermi level to be larger than $1/a_0$, with $a_0$ the nearest neighbour distance on the graphene lattice. This is satisfied
e.g. for conventional gold leads.
We thus obtain
\begin{eqnarray}
\lefteqn{G=\frac{(2\pi)^2}{\hbar}\frac{e^2}{\hbar^2}\frac{{\cal C}^2L^2}{N_L^2}\sum_{F F'}\sum_{bb'}\sum_{\sigma}
 \sum_{pp'}f_{Fbp}f_{Fbp'}f_{F'b'p}f_{F'b'p'}}\nonumber \\
&\times&\sum_{\vec R \vec R'}e^{i(F-F')(R_x-R'_x)}
|T_L(\vec R+\tau_p)|^2|T_R(\vec R' +\vec\tau_{p'})|^2 \nonumber \\
&\times&\int_{-\infty}^{\infty}d\omega \rho_L(\hbar\omega) \rho_R(\hbar\omega)\Biggl(- \frac{\partial n_F(\omega)}{\partial\omega}\Biggr) \nonumber
\\ && \times  G^{\rm ret}_{Fb\sigma}(\omega, x_L - x_R) G^{\rm adv}_{F'b'\sigma}(\omega, x_R - x_L),
\end{eqnarray}
where the constant $\cal C$ results from the integration over the $p_z$ orbitals.
Since we assume an extended tunneling region, the fast oscillating terms with $F=-F'$ are supposed to cancel. Furthermore,
we assume that both sublattices are equally coupled to the contacts, such that the sum over $\vec R$ $(\vec R')$ should give approximately the same results for $p=1$ and $p=2$. Hence, we can separate the sum over $p$ and $p'$ from the rest
and  exploit the relation
\begin{equation}
\sum_p f_{Fbp} f_{F b'p}= \delta_{b,b'}.
\end{equation}
Introducing now the transmission function
\begin{equation}
\label{transmission}
\Phi_L = \frac{{\cal C }L}{N_L} \sum_{\vec R,p} |T_L(\vec R+\tau_p)|^2 \rho_L(\varepsilon_F),
\end{equation}
and the analogous one, $\Phi_R$, for the right lead  we obtain Eq. (\ref{cond final}) when
%\begin{eqnarray}
%\lefteqn{G=\frac{(2\pi)^2}{\hbar}\frac{e^2}{\hbar^2}
%|\Phi_L|^2|\Phi_R|^2
%} \\
%&\times&\int_{-\infty}^{\infty}d\omega \Biggl(- \frac{\partial n_F(\omega)}{\partial\omega}\Biggr) \sum_{F b \sigma}| %G^{\rm ret}_{Fb\sigma}(\omega, x_L - x_R)|^2,\nonumber
%\end{eqnarray}
 the lead density of states is approximated with its value at the Fermi level (wide band
 approximation).
\section{The retarded Green's function}\label{Sec4:Greens funct}
As the essential information on the conductance is contained in
$G^{\rm ret }_{Fb\sigma}$, the main task lies in obtaining an
expression for this retarded Green's function in the case of a
toroidal SWNT.
\subsection{Time dependence}\label{time}
First we need to determine the time-dependence of the 1D electron
operator introduced in (\ref{el op}). To this extent, it is convenient to make use of the so called
bosonization identity, where the 1D  electron operator is written as a product of a fermionic operator, the Klein factor, and a bosonic field operator. A derivation of this identity can be found e.g. in \cite{Delft98}.  It yields
\begin{equation}
    \psi_{Fb\sigma}(x) = \eta_{Fb\sigma}e^{i\frac{2\pi}{L}\left(\mathcal N_{Fb\sigma} \mbox{\scriptsize sgn}(b) +\tilde{\phi}\right) x}e^{i\phi^\dagger_{Fb\sigma}(x)}e^{i\phi^{\phd}_{Fb\sigma}(x)},\label{psi boson}
\end{equation}
where the Klein factor $\eta_{Fb\sigma}$ lowers the number of electrons with given quantum numbers $Fb\sigma$ by one, and
\begin{equation}
 \label{phi}
    \phi^\dagger_{Fb\sigma}(x) = i\sum_{q>0}\frac{1}{\sqrt{|n_q|}} e^{-i\mbox{\scriptsize sgn}(b)
    qx}b^\dagger_{F\sigma \mbox{\scriptsize sgn}(b)q} e^{-\alpha q/2}.
\end{equation}
The infinitesimally small number $\alpha>0$ in (\ref{phi}) serves to avoid divergences due to the extension of the linearised dispersion relation.
Thus the time evolution  is determined by the
evolution of the Klein factors $\eta_{Fb\sigma}(t)$ and by
that of the bosonic fields $\phi_{Fb\sigma}(t)$ (\ref{phi}). Let
us then have a closer look at the Hamiltonian (\ref{H diag}). It
is made up of a part depending on the bosonic operators and one
depending on the number counting operators. These two types of
operators commute with each other. Therefore, the time evolution
of the operators $\eta_{Fb\sigma}(t)$ and $\phi_{Fb\sigma}(x,t)$ can
be treated separately. The bosonic part $\hat{H}_\mathrm{ring, b}$
yields the simple time evolution
$a^{\phd}_{\tilde{F}jq}(t) = e^{-\frac{i}{\hbar}\varepsilon_{\tilde F j q} t}a^{\phd}_{\tilde{F}jq}$, and analogously
for the operator
 $a^{\dagger}_{\tilde{F}jq}(t)$.
This can be related to the time evolution of the operators $b_{F\sigma q}$ appearing in the fields $\phi_{Fr\sigma}$.
The corresponding calculation for the Klein factors is a bit longer and yields
\begin{eqnarray}
 \eta_{Fb\sigma} (t) &=& e^{-\frac{i}{\hbar} \left[ \varepsilon_0 \left( \mathcal{N}_{Fb\sigma}
+ \frac{1}{2} + \tilde{\phi}\mbox{\scriptsize sgn}(b) \right) +
W_{00} \left( \mathcal{N} +\frac{1}{2} \right)  -\mu_g \right] t}
\eta_{Fb\sigma} \nonumber
\\
&=:& Y_{Fb\sigma } (t) \eta_{Fb\sigma}(0), \label{Y}
\end{eqnarray}
as well as
\begin{eqnarray}
 \eta^\dagger_{Fb\sigma} (t) &=&
e^{\frac{i}{\hbar} \left[ \varepsilon_0 \left( \mathcal{N}_{Fb\sigma} - \frac{1}{2} + \tilde{\phi}\mbox{\scriptsize sgn}(b) \right) + W_{00} \left( \mathcal{N} -\frac{1}{2} \right)  -\mu_g \right] t} \eta_{Fb\sigma} \nonumber
\\
&=:& Y^+_{Fb\sigma } (t) \eta^\dagger_{Fb\sigma}(0). \label{Y+}
\end{eqnarray}
\subsection{Separation of the correlation function}
Let us now turn to the averaging process. As the number of electrons $\mathcal{N}$ in the dot can vary we have to work in the grand canonical ensemble,
 summing over all values of $\mathcal{N}$. The correlation function
\begin{eqnarray}
&&\left\langle  \psi^\dagger_{F b\sigma}(x',
0)\psi^{\phd}_{Fb\sigma}(x, t)\right\rangle = \nonumber
\\
&& = \mbox{Tr}\left\{ \hat{\rho}_{\mathrm{ring},\mathcal{N}}
\psi^\dagger_{Fb\sigma}(x', 0) \psi^{\phd}_{Fb\sigma}(x, t)
\right\},
\end{eqnarray}
is a thermal expectation value
with respect to the grandcanonical density matrix operator
%\begin{eqnarray}
 $\hat{\rho}_{\mathrm{ring},\mathcal{N}} = \frac{1}{Z}e^{-\beta
\hat{H}_\mathrm{ring} }$,
and the trace is over the many body states $\left|\vec{N}, \vec{m}\right\rangle$, cf. (\ref{eigenstates}),
where
$\vec{N}$ and $\vec{m}$
%&=&(N_{K_0 r\uparrow}, N_{K_0 r\downarrow}, N_{K_0 l\uparrow}, N_{K_0 l\downarrow}, \nonumber
%\\ && N_{-K_0 r\uparrow}, N_{-K_0 r\downarrow}, N_{-K_0 l\uparrow}, N_{-K_0 l\downarrow}).
%\end{eqnarray}
define the fermionic and bosonic configurations, respectively.
%For each value of $\vec{N}$, one has to include  the  states which contain
%the bosonic excitations of the interacting electrons given by
%\begin{eqnarray}
%\left \{ \prod_{ \tilde{F}j,q \neq 0,} \frac{(a^\dagger_{\tilde{F} j q})^{m_{\tilde{F} jq}}}{\sqrt{m_{\tilde{F} jq}!}} %\left|\vec{N}\right\rangle \right \}   =: \left \{\left|\vec{N}, \vec{m}\right\rangle \right \}.
%\label{eigenstates}
%\end{eqnarray}
%The parameter $m_{\tilde{F} jq}$ counts the number of bosonic excitations in the channel $\tilde F j$ with momentum $q$.
Accordingly, the trace is
\begin{eqnarray}
\mbox{Tr}\left\{\ldots  \right\} \equiv \sum_{\mathcal{N}} \sum_{\{\vec{N}\}_{\mathcal{N}}} \sum_{\vec{m}}\left\langle \vec{N}, \vec{m}\right|\ldots \left|\vec{N}, \vec{m}\right\rangle.
\end{eqnarray}
%Inserting the correlation function into the trace yields
%\begin{eqnarray}
%&& \left\langle  \psi^\dagger_{Fb\sigma'}(x',
%0)\psi^{\phd}_{Fb\sigma}(x, t)\right\rangle = \sum_{\mathcal{N}}
%\sum_{\{\vec{N}\}_{\mathcal{N}}} \sum_{\vec{m}} \times \nonumber
%\\
%&& \times \left\langle \vec{N}, \vec{m}\right|
%\hat{\rho}_\mathrm{ring}  \psi^\dagger_{Fb\sigma'}(x', 0)
%\psi^{\phd}_{Fb\sigma}(x, t)\left|\vec{N}, \vec{m}\right\rangle.
%\nonumber
%\\
%\end{eqnarray}
In order to make use of the unitarity of the Klein factors, we
exploit the fact that the above correlator only depends  on the
time difference.
%We can therefore evaluate as well $\left\langle
%\psi^\dagger_{Fb\sigma}(x', -t)\psi^{\phd}_{Fb\sigma}(x,
%0)\right\rangle$.
Thus,
 with the help of the definitions (\ref{psi boson}), (\ref{phi}) and (\ref{Y+})  we get
%\begin{eqnarray*}
%&& \left\langle  \psi^\dagger_{Fb\sigma}(x',
%0)\psi^{\phd}_{Fb\sigma}(x, t)\right\rangle =
%\delta_{FF'}\delta_{rr'}\delta_{\sigma\sigma'}
%\sum_{\mathcal{N}} \sum_{\{\vec{N}\}_{\mathcal{N}}} \sum_{\vec{m}}
 %\nonumber
%\\
%&&\times\left\langle \vec{N}, \vec{m}\right|
%\hat{\rho}_\mathrm{ring} e^{-i\phi^\dagger_{Fb\sigma}(x',
%0)}e^{-i\phi^{\phd}_{Fb\sigma}(x', 0)}  \times \nonumber
%\\
%&& \times e^{-i\frac{2\pi}{L}\left(\mathcal N^{\phd}_{Fb\sigma}
%\mbox{\scriptsize sgn}(b) +\tilde{\phi}\right)
%x'}\eta^\dagger_{Fb\sigma'}(0)
%Y^{\phd}_{Fb\sigma}(t)\eta^{\phd}_{Fb\sigma}(0) \times \nonumber
%\\
%&& \times e^{i\frac{2\pi}{L}\left(\mathcal N^{\phd}_{Fb\sigma}
%\mbox{\scriptsize sgn}(r) +\tilde{\phi}\right) x}
%e^{i\phi^\dagger_{Fb\sigma}(x, t)}e^{i\phi^{\phd}_{Fb\sigma}(x,
%t)}
% \left|\vec{N}, \vec{m}\right\rangle.
%\end{eqnarray*}
%
%Further, since only the Klein factors can change
%the number of electrons in this expression, we only need to take
%into account the case $F=F'$, $r=r'$, $\sigma=\sigma'$ as
%otherwise we would get scalar products of states from different
%Hilbert spaces yielding no contribution.
%
\begin{eqnarray}
\lefteqn{\left\langle  \psi^\dagger_{Fb\sigma}(x',0)\psi^{\phd}_{Fb\sigma}(x, t)\right\rangle =\left\langle  \psi^\dagger_{Fb\sigma}(x',
-t)\psi^{\phd}_{Fb\sigma}(x, 0)\right\rangle } \nonumber
\\
&=&  \sum_{\mathcal{N}} \sum_{\{\vec{N}\}_{\mathcal{N}}} \sum_{\vec{m}} \left\langle \vec{N}, \vec{m}\right| \hat{\rho}_\mathrm{ring}  e^{-i\phi^\dagger_{Fb\sigma}(x', -t)}e^{-i\phi^{\phd}_{Fb\sigma}(x', -t)} \nonumber
\\
&\times& e^{-i\frac{2\pi}{L}\left(\mathcal N^{\phd}_{Fb\sigma} \mbox{\scriptsize sgn}(b) +\tilde{\phi}\right) x'} Y^+_{Fb\sigma}(-t) \underbrace{\eta^\dagger_{Fb\sigma}(0) \eta^{\phd}_{Fb\sigma}(0)}_{= 1} \nonumber
\\
 & \times & e^{i\frac{2\pi}{L}\left(\mathcal N^{\phd}_{Fb\sigma} \mbox{\scriptsize sgn}(b) +\tilde{\phi}\right) x} e^{i\phi^\dagger_{Fb\sigma}(x, 0)}e^{i\phi^{\phd}_{Fb\sigma}(x, 0)}
 \left|\vec{N}, \vec{m}\right\rangle.
\end{eqnarray}
As the bosonic and the fermionic operators commute, the correlation function can now be separated into a fermionic and a bosonic part:
\begin{eqnarray}
&& \left\langle  \psi^\dagger_{Fb\sigma}(x', -t)\psi^{\phd}_{Fb\sigma}(x, 0)\right\rangle = \nonumber
\\
&& \left\langle  \psi^\dagger_{Fb\sigma}(x', -t) \psi^{\phd}_{Fb\sigma}(x, 0)\right\rangle_f \left\langle  \psi^\dagger_{Fb\sigma}(x', -t)\psi^{\phd}_{Fb\sigma}(x, 0)\right\rangle_b, \nonumber
\end{eqnarray}
where, with $E_{\rm ring,f}(\vec N)$ being the eigenvalues of $\hat H_{\rm ring,f}$,
\begin{eqnarray}
 %\lefteqn{
 &&\hspace{-0.5cm}\left\langle  \psi^\dagger_{Fb\sigma}(x', -t) \psi^{\phd}_{Fb\sigma}(x, 0)\right\rangle_f  =: \sum_{\mathcal{N}} \sum_{\{\vec{N}\}_{\mathcal{N}}} \frac{1}{Z_f}e^{-\beta E^{\phd}_{\mathrm{ring, f}} }
 %}
 \nonumber\\
& \times & Y^+_{Fb\sigma}(-t) e^{-i\frac{2\pi}{L}\left( N_{Fb\sigma} \mbox{\scriptsize sgn}(b) +\tilde{\phi}\right) (x'-x)}
% e^{i\frac{2\pi}{L}\left(\mathcal N_{Fb\sigma} \mbox{\scriptsize sgn}(b) +\tilde{\phi}\right) x}
, \label{green-fermionic-1}
\end{eqnarray}
and
\begin{eqnarray}
 \lefteqn{\left\langle  \psi^\dagger_{Fb\sigma}(x', -t) \psi^{\phd}_{Fb\sigma}(x, 0)\right\rangle_b =: \frac{1}{Z_b}\sum_{\vec{m}} \left\langle \vec{0}, \vec{m}\right|e^{-\beta \hat{H}^{\phd}_{\mathrm{ring, b}}} } \nonumber
\\
& \times & e^{-i\phi^\dagger_{Fb\sigma}(x', -t)}e^{-i\phi^{\phd}_{Fb\sigma}(x',- t)}e^{i\phi^\dagger_{Fb\sigma}(x, 0)}e^{i\phi^{\phd}_{Fb\sigma}(x, 0)} \left|\vec{0}, \vec{m}\right\rangle. \nonumber
\\
&& \label{c6 sep}
\end{eqnarray}
In the bosonic part we only need to take care of the state $\left|\vec{0}, \vec{m}\right\rangle$, as the bosonic excitation spectrum is the same for all charge states.
The partition functions read
\begin{eqnarray}
Z_f &=&\sum_{\mathcal{N}} \sum_{\{\vec{N}\}_{\mathcal{N}}}  \left\langle \vec{N}, 0 \right| e^{-\beta \hat{H}_{\mathrm{ring,f }}} \left|\vec{N}, 0\right\rangle ,
\\
Z_b &=& \sum_{\vec{m}} \left\langle \vec{0}, \vec{m}\right|e^{-\beta \hat{H}_{\mathrm{ring, b}}} \left|\vec{0}, \vec{m}\right\rangle.
\end{eqnarray}
In order to evaluate the Green's function we also need the correlator $\left \langle \psi_{F b \sigma} (x,t) \psi^\dagger_{F b \sigma}(x',0)\right\rangle$. As above, it can be factorized in a  fermionic and a bosonic
part. It holds,
\begin{equation}
\left \langle \psi^\dagger_{F b \sigma} (x',-t) \psi_{F b \sigma}(x,0)\right\rangle_b =
\left \langle \psi_{F b \sigma} (x,t) \psi^\dagger_{F b \sigma}(x',0)\right\rangle_b^*.
\end{equation}
For the fermionic part we find
\begin{eqnarray}
 %\lefteqn{
 &&\hspace{-0.5cm}\left\langle  \psi^{\phd}_{Fb\sigma}(x, t) \psi^\dagger_{Fb\sigma}(x', 0)\right\rangle_f  =: \sum_{\mathcal{N}} \sum_{\{\vec{N}\}_{\mathcal{N}}} \frac{1}{Z_f}e^{-\beta E^{\phd}_{\mathrm{ring, f}} }
 %}
 \nonumber\\
& \times & Y_{Fb\sigma}(t) e^{-i\frac{2\pi}{L}\left( N_{Fb\sigma} \mbox{\scriptsize sgn}(b) +\tilde{\phi}\right) (x'-x)}
% e^{i\frac{2\pi}{L}\left(\mathcal N_{Fb\sigma} \mbox{\scriptsize sgn}(b) +\tilde{\phi}\right) x}
. \label{green-fermionic-2}
\end{eqnarray}

\subsection{The bosonic part of the Green's function}
The evaluation of the bosonic part of the Green's function is lengthy but standard.
Following e.g. \cite{Delft98} we obtain
\begin{eqnarray}
\lefteqn{\left\langle  \psi^\dagger_{Fb\sigma}(x', -t) \psi^{\phd}_{Fb\sigma}(x, 0)\right\rangle_b=
\frac{1}{1-e^{-\alpha\frac{2\pi}{L}}}} \label{green-bosonic}\\
&\times&\exp\left\{
-\frac{1}{4}\sum_{q>0}\frac{e^{-\alpha q}}{n_q} \right.\left[ \sum_{\tilde F j}Q_{\tilde F j q}(-t, {\rm sgn}(b)(x'-x))\right.
\nonumber\\
&+& \left.
2C^2_{\oplus cqq}[Q_{ c q}(-t, (x'-x))+Q_{ c q}(-t, -(x'-x))]
\right]\left.\right\},\nonumber
\end{eqnarray}
where we introduced the correlation function
\begin{eqnarray}
Q_{\gamma q}(t,x)&:=&i\sin \Bigl( \frac{\varepsilon_{\gamma q} t}{\hbar} -q x\Bigr)
 \\ &+&
\left[1-\cos\Bigl(\frac{\varepsilon_{\gamma q} t}{\hbar}-qx\Bigr)\right]
\coth\Bigl(\frac{\beta\varepsilon_{\gamma q} }{2}\Bigr).\nonumber
\end{eqnarray}
The parameters $C^2_{\oplus cqq}$ entering in the Bogoliubov transformation which diagonalizes the bosonic part of
the Hamiltonian, cf. (\ref{Ham}),  are defined in the Appendix A.
To proceed in the analytic evaluation of  the bosonic part of the Green's function we perform the following
two approximations. The first one consists in linearising the energies $\varepsilon_{cq}$ which belong to the diagonalised Hamiltonian (\ref{Ham}):
\begin{eqnarray}
\varepsilon_{cq} &=&  \hbar v_c |q| = \hbar \frac{v_F}{K_c} |q|,\label{c6 lin en}
\end{eqnarray}
where we introduced the charge velocity $v_c$. The coupling constant $K_c$ is given by
%\begin{eqnarray}
$ K_c = \Bigl(\sqrt{1+\frac{8W_{q_0 -q_0}}{\varepsilon_0}}\Bigr)^{-1/2},\ \ q_0 = 2\pi/L$,
%\end{eqnarray}
and is determined by the interaction for small values of the momentum $q$. In case of a repulsive interaction $K_c$ is always smaller than 1, yielding $v_c>v_F$, $\varepsilon_{cq}>\varepsilon_{0q}$ and $\varepsilon_c=\varepsilon_0/K_c>\varepsilon_0$.
\\
The second approximation refers to the parameter $C^2_{\oplus cqq}$.  We already assumed in section \ref{Int Ham} that, due to the long range character of the interaction, a momentum cut-off $q_c$ can be introduced. For high values of the momentum $q$, the parameter $C^2_{\oplus cqq}$, which depends on the interaction, has to vanish. We assume an exponential decay in $q$-space,
$
C^2_{\oplus cqq} = C^2_{\oplus cq_0}e^{-q/q_c}.
%\label{c6 decay}
$
The relation between $C^2_{\oplus cq_0}$ and $K_c$ is found with the help of the transformation parameters $\alpha_q$ and $\beta_q$:
\begin{eqnarray*}
\alpha_q = \frac{1}{2}\sqrt{\frac{\varepsilon_{0q}}{2 \varepsilon_{cq}}},\ \ \ \beta_q = \frac{1}{2}\sqrt{\frac{\varepsilon_{cq}}{2 \epsilon_{0q}}}.
\end{eqnarray*}
For $q= q_0$ they read:
%\begin{eqnarray*}
$\alpha_{q_0} = \frac{1}{2}\sqrt{K_c/2},\ \ \ \beta_{q_0} = 1/(2\sqrt{2K_c})$,
%\end{eqnarray*}
leading to the expression
\begin{eqnarray}
C^2_{\oplus cq_0} = \frac{1}{64}\left( \frac{1}{\beta_{q_0}}-\frac{1}{\alpha_{q_0}}\right)^2 =  \frac{1}{8}\left( K_c^{1/2}-K_c^{-1/2}\right)^2.
\end{eqnarray}
With these two approximations we can evaluate the bosonic correlation function at low
 temperatures, $k_B T \ll \varepsilon_0$ (in this regime $\coth(\beta\varepsilon_0/2)\approx 1$). We find, see
 Eq.(\ref{appD bos}) in Appendix \ref{appB},
 \begin{eqnarray}
\lefteqn{\hspace{-1cm}\left \langle \psi^\dagger_{F b \sigma} (x',-t) \psi_{F b \sigma}(x,0)\right\rangle_b = \left[ 1-z\left(\alpha + \frac{1}{q_c}, 0\right)\right]^{C^2_{\oplus cq_0}}} \nonumber
\\
&\times &\sum_{k=0}^\infty \sum_{l=0}^\infty F_{k,l}(b,x'-x)
e^{\frac{i}{\hbar}\varepsilon_0 kt}e^{\frac{i}{\hbar}\varepsilon_c l t},
\label{boson-corr-1}
\end{eqnarray}
where $z(\alpha, x) = e^{-\alpha\frac{2\pi}{L}}e^{i\frac{2\pi}{L}x}$.
\subsection{The Fourier transform of the Green's function}
Despite its intricate form, see (\ref{appD bos}), the time dependence of the bosonic part of the Green's function is trivial.
Gathering then together bosonic and fermionic contributions we obtain for the Fourier transform of the
SWNT Green's function
\begin{eqnarray}
&&G^{\rm ret}_{Fb\sigma} (\omega, x' - x) = \left[ 1-z\left(\alpha + \frac{1}{q_c}, 0\right)\right]^{C^2_{\oplus cq_0}}
\nonumber
\\
&\times &
\sum_{\mathcal{N}} \sum_{\{\vec{N}\}_{\mathcal{N}}} \frac{1}{Z_f}e^{-\beta E^{\phd}_{\mathrm{ring, f}} }
e^{-i\frac{2\pi}{L}( N_{Fb\sigma}{\rm sgn}(b)+ \tilde\phi)(x'-x)}
\nonumber\\&\times&
\sum_{k,l=0}^\infty \left[\frac{F_{k,l}(b,x'-x)}{
\hbar\omega-{\cal E}_{kl}({ N}_{Fb\sigma},{ N},\mu_g,\tilde\phi)+i\eta
}\right. \label{Green result}\\&+&
\left.
\frac{F^*_{k,l}(b,x'-x)}{
\hbar\omega-{\cal E}_{-k-l}({N}_{Fb\sigma}+1,{ N}+1,\mu_g,\tilde\phi)+i\eta
}
\right]\nonumber
,
\end{eqnarray}
where $\eta$ is a positive infinitesimal and  the energy
\begin{eqnarray}
&&{\cal E}_{-k-l}({ N}_{Fb\sigma}+1,{\cal N}+1,\mu_g,\tilde\phi)=-\mu_g+ \label{pole-energies} \\
&&\varepsilon_0( N_{Fb\sigma}+k+\frac{1}{2}+\tilde\phi{\rm sgn}(b) )  + W_{00} (\mathcal N + \frac{1}{2})   +\varepsilon_c l \nonumber
\end{eqnarray}
is  the energy needed to add a particle in the branch $(Fb\sigma)$ to a system with ${\mathcal N}$ electrons together with a difference of  $k$ neutral and $l$ charged bosonic excitations.
Likewise, $-{\cal E}_{kl}({ N}_{Fb\sigma},N_c,\mu_g,\tilde\phi)$ is the energy gained by removing a particle on
branch $(Fb\sigma)$ from a system with $\mathcal N$ electrons plus bosonic excitations.
\section{Results for
the conductance} \label{Sec5:results cond}
We are now finally at a stage in which we can discuss some features of the conductance of a toroidal SWNT,
 as they follow from the conductance formula (\ref{cond final}) together with the expression for the
 Fourier transform of the Green's function (\ref{Green result}).

\subsection{Summation over the relevant configurations}
The expression (\ref{Green result}) is quite intricate, as it requires a summation over all the fermionic and bosonic
configurations. A noticeable simplification, however, occurs in the  low temperature regime $\varepsilon_0>k_B T$,
which we shall address in the following. In this regime in fact we need to consider \emph{ground state configurations only} of the SWNT ring, i.e., we can neglect bosonic and fermionic excitations. Neglecting bosonic excitations means to
 set $k=l=0$ in the sum in (\ref{Green result}), such that the only effect of the bosonic part of the Green's function is in the interaction-dependent prefactor $\left[ 1-z\left(\alpha + \frac{1}{q_c}, 0\right)\right]^{C^2_{\oplus cq_0}}$.
   With $\tilde\alpha= \alpha +q_c^{-1}$ it yields a conductance suppression proportional to $(\tilde\alpha/L)^{2C^2_{\oplus cq_0}}=(\tilde\alpha/L)^{(K_c+K_c^{-1}-2)/4}$
with respect to the non-interacting case. To find the relevant
 fermionic configurations we observe that at low temperatures the contribution to the trace comes from those values of the total electron number $\mathcal N$ and of the total current $\mathcal J$ which minimize the energy $E_{\mathrm{ring, f}}({\vec N})$.
   This energy is provided by the fermionic part of the Hamiltonian (\ref{H diag}) and is determined by the applied gate voltage and magnetic field.
 \subsubsection{Spinless case}
\begin{figure}[t]
 \includegraphics{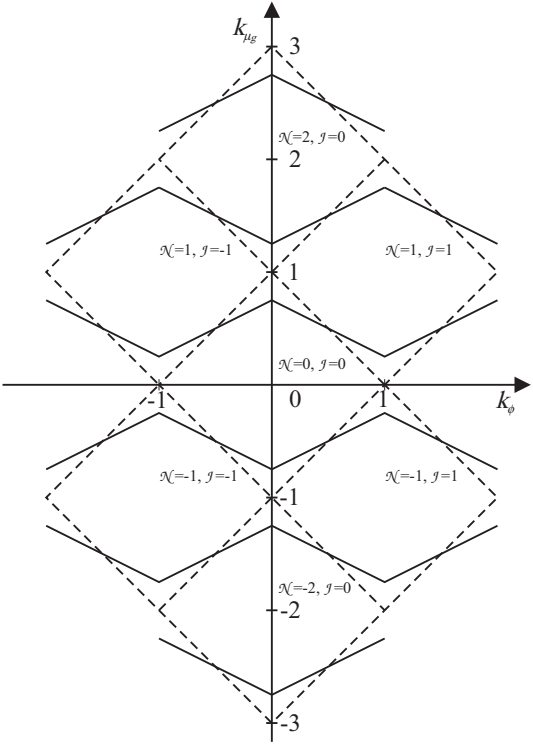}
% If not, use
%\vspace{5cm}       % Give the correct figure height in cm
\caption{Ground states resonances for a ring with two fermionic species $b=R,L$:
Along the lines the energies are degenerate and transport is there maximal. Continuous lines correspond to the interaction parameter $\tilde W = 2$, dotted lines  to the non-interacting case, $\tilde W = 1$.}
\label{fig:1}       % Give a unique label
\end{figure}

\begin{figure}[t]
  \includegraphics{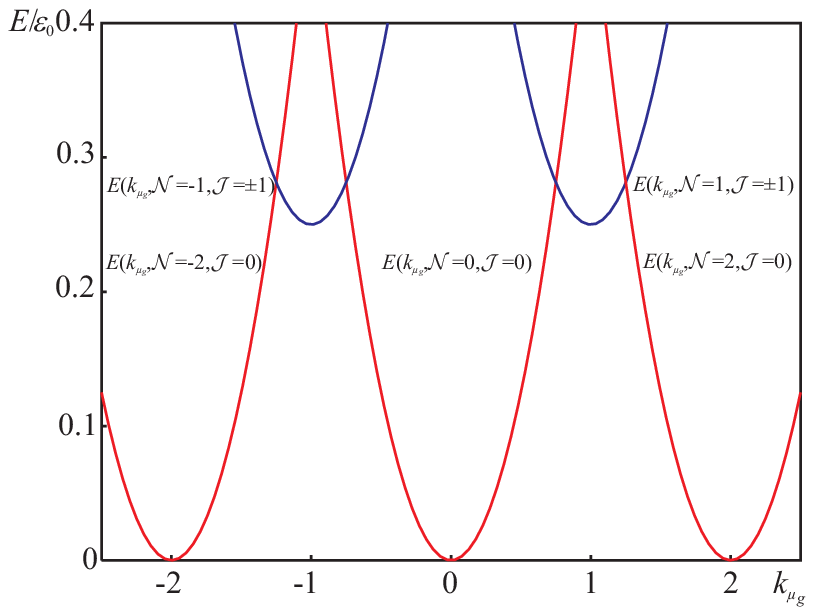}
%
% If not, use
%\vspace{5cm}       % Give the correct figure height in cm
\caption{Ground state energies $E(k_{\mu_g}, \mathcal N, \mathcal J)$ for $k_\phi=0$ and interaction parameter
$\tilde W=2$. Notice that the parabolas corresponding to an odd electron number are doubly degenerate.}
\label{fig:2}       % Give a unique label
\end{figure}

 To clarify the ideas, let us first consider a ring with only two species of spinless electrons characterized by $b=R,L$. Then, $\hat{H}_{\mathrm{ring,f }}$ simplifies to
\begin{eqnarray}
\hat{H}_{\mathrm{ring, f}}&=&  \frac{\varepsilon_0}{4} \left[\tilde{W} ({\mathcal N}-k_{\mu_g})^2 +
({\mathcal J}-k_\phi)^2 \right],
\end{eqnarray}
where now $k_{\mu_g} = \frac{\mu_g}{W_{00} + \varepsilon_0 /2}$, $k_\phi = -2 \tilde \phi$ and $\tilde{W} = 1+ \frac{2W_{00}}{\varepsilon_0}$ hold. Figure \ref{fig:1} shows which ground state $(\mathcal N, \mathcal J)$ is occupied in the $(k_\phi,k_{\mu_g})$-plane. The continuous lines are obtained by setting $\tilde W = 2$, whereas the dotted ones belong to the non-interacting case $\tilde W = 1$. Along such lines the energies of neighboring ground states are degenerate. At zero temperature these are the conductance resonances.
Figure \ref{fig:2} shows the ground state energies $E(k_{\mu_g}, \mathcal N, \mathcal J)$ for $k_\phi=0$.
For values of $k_{\mu_g}$ around zero the ring is in the state $(\mathcal N = 0, \mathcal J=0)$. Raising the external gate voltage above $k_{\mu_g}=0.75$  leads to a transition to a charged state with one electron. Above $k_{\mu_g} = 1.25$ the state $(\mathcal N = 2, \mathcal J=0)$ is the one with the lowest energy. We observe that along $k_\phi = 0$ for the states with an odd number of electrons there is also a degeneracy between the states with the same charge but different current
 directions.
\subsubsection{SWNT ring}
Let us now turn back to the SWNT ring. Due to the presence of the spin and orbital degrees of freedom, the fermionic part of the Hamiltonian is quadratic in all of the
variables ${\cal N}_{\tilde F j}$ and ${\cal J}_{\tilde F j}$, cf. (\ref{H diag}).
However, it remains gapped only with respect to the total charge and current operators. Let us then
introduce for given $\mu_g$ and $\tilde\phi$ the integer numbers $N_c$ and $J_c$
which minimize $E_{\rm ring,f}$.
Outside resonance, only one couple  $(N_c,J_c)$ contributes to the grandcanonical sum in (\ref{Green result}).
 Nearby resonance  we have to include states which differ by one electron and with corresponding current configurations differing by plus/minus one.
 %characterized by $N_c$, $J_c$ and $N_c+ |\hat e_{F'b'\sigma'}|$, $J_c+{\rm sgn}b'$.
 Moreover, in the low temperature regime $k_BT<\varepsilon_0$ the quadratic forms in $\hat H_{\rm ring,f}$  impose that we only have to take into account those configurations $\{ \vec N\}$, where the number of electrons in each configuration $(F'b'\sigma')$ differs at most by one (i.e., we neglect "fermionic excitations").
 From now on we measure the number of excess electrons or holes from the charge neutrality point where ${\vec N}=0$.
 Thus, ${ N}_{F'b'\sigma'}$ can only assume the values $m$ and $m+1$, where $m\ge 0$ above and $m<0$ below the
 charge neutrality point. This means that $N_c$, obtained by  summing over the occupation of all the branches  $(F'b'\sigma')$, can be written as
%\begin{equation}
$N_c=8m+l,\quad m \in Z, \quad l=0,1,\ldots 7.$
%\label{Nc}
%\end{equation}
Notice that this implies an eight-electron periodicity as a function of the gate voltage.
In contrast, as $J_c$ counts the occupation difference between the orbital branches, $b=R,L$, it can only take values
between $-4$ and $+4$.
%With these considerations one finds that $P(\vec{N},N_c,J_c)$ is \emph{independent of the specific configuration} ${\vec N}$ and reads
%\begin{equation}
%P({\vec N},N_c,J_c)&:=&P(N_c,J_c)=\frac{1}{\sum_{\{{\vec N} \}_{N_c,J_c}}}.
%\end{equation}
%Accordingly, the configuration sum in  Eq. (\ref{green-fermionic-3}) can be explicitly performed.
We can thus simplify the trace operation considerably by summing only over the dominant configurations in (\ref{Green result}).
These are the configurations where $N_c\to N_c+1$ with  ${ N}_{Fb\sigma}=m$, and $N_c+1\to N_c$
with  ${ N}_{Fb\sigma}=m+1$. We obtain
 %\begin{widetext}
\begin{eqnarray}
&&G^{\rm ret}_{Fb\sigma} (\omega, x' - x) = \left[ 1-z\left(\alpha + \frac{1}{q_c}, 0\right)\right]^{C^2_{\oplus cq_0}}
  \nonumber
\\
&\times&
\frac{e^{-\beta E^{\phd}_{\mathrm{ring, f}}(\vec N|_{N_c+1} }+e^{-\beta E^{\phd}_{\mathrm{ring, f}}(\vec N|_{N_c+1} )}
e^{-i\frac{2\pi}{L}{\rm sgn}(b)(x'-x)}}{
e^{-\beta E^{\phd}_{\mathrm{ring, f}}(\vec N|_{N_c}) }+e^{-\beta E^{\phd}_{\mathrm{ring, f}}(\vec N|_{N_c+1} )
}} \nonumber \\
&\times & \hspace{-0.1cm}
 \frac{e^{-i\frac{2\pi}{L}(m{\rm sgn}(b)+\tilde\phi)(x'-x)}}{\hbar\omega-{\cal E}_{00}(m+1,N_c+1,\mu_g,\tilde\phi)+i\eta}.
\label{Green result-2}
\end{eqnarray}
%\end{widetext}
%
Recalling that
$E^{\phd}_{\mathrm{ring, f}}(\vec N|_{N_c+1})-E^{\phd}_{\mathrm{ring, f}}(\vec N|_{N_c+1})={\cal E}_{00}(m+1,N_c+1,\mu_g,\tilde\phi)\equiv {\cal E}_{0}$, we obtain
\begin{eqnarray}
&&|G^{\rm ret}_{Fb\sigma} (\omega, x' - x)|^2 = \left[ 1-z\left(\alpha + \frac{1}{q_c}, 0\right)\right]^{2C^2_{\oplus cq_0}}
  \nonumber
\\
&\times & \hspace{-0.1cm}
 \frac{\cos^2(\Delta x)+\sin^2 (\Delta x)\tanh^2\Bigl(\frac{\beta{\cal E}_0}{2}\Bigr)}{|\hbar\omega-{\cal E}_{00}(m+1,N_c+1,\mu_g,\tilde\phi)+i\eta|^2},
\label{Green result-3}
\end{eqnarray}
where $\Delta x:=\frac{\pi}{L}(m{\rm sgn}(b)(x'-x) $.
When Eq. (\ref{Green result-3}) is inserted in  Eq. (\ref{cond final}) the conductance diverges, as a consequence of the
 presence of the infinitesimal convergence factor $i\eta$ rather than of a finite line-width \cite{Kinaret98,Rollbühler99}. Nevertheless, Eq. (\ref{Green result-3})
 allows already to discuss the main qualitative features of the conductance, namely the resonance pattern as a function
 of the applied gate voltage and magnetic field.

 To be definite, conductance resonances occur at a degeneracy for the removal of either a counter-clockwise or a clockwise propagating electron. Moreover, as in the spinless case, double resonances occur for special  values of the applied magnetic flux and gate voltage, such that clockwise and anti-clockwise propagating modes are simultaneously at resonance. As the ring energy is independent of the spin and Fermi number degrees of freedom,
a double resonance is uniquely fixed by $N_c$, $J_c$. Let us consider e.g. $N_c=0$, $J_c=0$ and $N_c+1=1$, $J_c\pm 1$.
Then it only exists a configuration with $N_c=0$, while there are four equivalent configurations with $N_c=+1$, $J_c=1$
and four with $N_c=1$, $J_c=-1$ contributing to the double resonance.

\begin{figure}[t]
    \centering
        \includegraphics{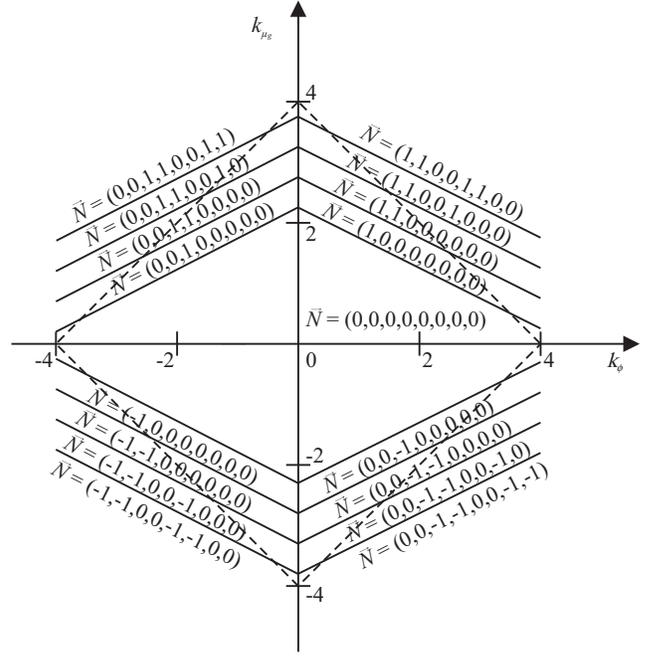}
    \caption{Conductance resonances of the carbon nanotube ring as a function of $k_{\mu_g}$ and $k_\phi$. The interaction is chosen to be $\tilde W=2$. The lines with different slopes correspond to an excess of electrons propagating
    in opposite directions (counterclockwise and clockwise). Dashed lines are for the non interacting case.   }
        \label{Frsigma}
\end{figure}

The resonant  pattern is easily obtained by approximating the derivative of the
Fermi function in Eq. (\ref{cond final}) with a delta function \cite{Kinaret98} such that $G\propto \sum_{Fb\sigma} |G^{\rm ret}_{Fb\sigma}(0,x'-x)|^2$, and then evaluating the poles of (\ref{Green result-3}) as a function of $k_{\mu_g}$ and
 $k_\phi$. Upon recalling that $N_c=8m+l$, with $m \in Z$ and $l=0, 1, ..7$, we find
 \begin{equation}
 k_{\mu_g}= 8m + \frac{4}{\tilde W}\Bigl(1+\frac{W_{00}}{\varepsilon_0} \Bigr) +\frac{8W_{00}}{\varepsilon_0\tilde W}l
 -\frac{k_{\phi}}{\tilde W}{\rm sgn} (b)\; .
 \end{equation}
 Notice that this implies a distance between adjacent resonant lines, with $l$ and $l+1$, of $\Delta k_{\mu_g}=\frac{8W_{00}}{\varepsilon_0\tilde W}$.
 The corresponding results are  shown in figure \ref{Frsigma}.
We choose a value of $\tilde W = 2$ for the plot. The solutions for the non-interacting system are  shown with dotted lines. In the non-interacting case the ring is occupied by a multiple of four electrons. From one square to the next other four electrons can be accomodated at one time, as there is no charging energy. In the presence of interactions, however, the picture is entirely different. The former squares are deformed to hexagons which only touch at their vertical boundary. There, the two ground states are degenerate. The conductance is zero here. In the parallel stripes in figure \ref{Frsigma} we indicated one possible fermionic configuration only. There is no difference in the energy if we add the first electron to a spin-down state instead of a spin-up one as we did. The choice of $F$ is arbitrary as well. The main difference to the non-interacting case is that, similar to the spinless case, for certain values of $k_{\mu_g}$ no conductance resonances can be achieved.

Finally, Eq. (\ref{Green result-3}) predicts a periodic pattern of the conductance as a function of the distance between left and right leads, with maxima when the distance between leads is $x_R-x_L=L/2$, cf. Fig. \ref{ring}.
\section{Conclusions}\label{conc}
In this work the linear conductance of an interacting toroidal SWNT weakly coupled to leads and pierced by a magnetic field  has been analyzed. We focussed on low temperatures $k_B T <\varepsilon_0$, where $\varepsilon_0$ is the mean level spacing. In this regime Coulomb blockade effects are known to crucially affect the conductance of a straight SWNT as a function of an applied gate voltage \cite{Sapmaz05,Moriyama05}. Likewise Coulomb interactions strongly influence the
conductance characteristics of the ring. Due to charging effects, the electrons can enter the ring only one by one and not up to four at the time as in the non-interacting case. While in straight tubes the conductance pattern has a four-electron periodicity, for toroidal SWNTs we
 predict a periodicity of eight electrons. The conductance exhibits Coulomb resonances, and the resonance pattern is a function of the applied magnetic field $\phi$ and gate voltage $\mu_g$. In contrast to  the non-interacting case, where the  parameters $\phi$ and ${\mu_g}$ can always be tuned to match a resonance condition,
in the interacting system this is no longer possible and we find that at certain values of $\mu_g$ a window opens where no conductance resonances can be obtained. Interference effects are manifested in the occurrence of double resonances where electron trajectories propagating clockwise and anticlockwise are degenerate.
Finally,   the conductance as a function of the distance between left and right leads is  maximal when the distance between leads is half of the circumference length, i.e., when clockwise and anti-clockwise propagating electrons have to
cover the same source to drain distance.

The conductance of toroidal SWTNs has already been measured in the experiments \cite{Shea00}, though not in the
Coulomb blockade regime. Thus, we believe that the possible verification of our predictions is within the reach of present experiments.
\appendix\label{appA}
\section{Bogoliubov transformation}
The non-diagonal part of (\ref{V boson sc}) is given by
\begin{eqnarray}
&& \hat{H}_{\mathrm{nd}} =  \varepsilon_{0}\sum_{q \neq 0}|n_q|  b_{\oplus cq}^\dagger b_{\oplus cq} + \frac{1}{2} \sum_{q \neq0} 4 |n_q|W_{q,-q} \times\nonumber
\\
&\times&  \Bigl(b_{\oplus cq} b_{\oplus c -q} + b_{\oplus cq} b^\dagger_{\oplus cq} + b^\dagger_{\oplus c-q} b_{\oplus c-q} + b^\dagger_{\oplus c-q} b^\dagger_{\oplus cq}\Bigr). \nonumber
\\ \label{appB H not diag}
\end{eqnarray}
The goal of the Bogoliubov transformation is the derivation of new operators, in terms of which $\hat{H}_{\mathrm{nd}}$ is diagonal. In general, we can write them as a linear combination of the old bosonic operators $b^\dagger_{\oplus cq}$ and $b_{\oplus cq}$:
\begin{eqnarray}
a^\dagger_{\oplus cq} &=& \sum_{q'} (U_{qq'} b^\dagger_{\oplus cq'} - V_{qq'} b_{\oplus cq'}), \label{appB adagger}
\\ \label{appB a}
a_{\oplus cq} &=& \sum_{q'} ( - V_{qq'} b^\dagger_{\oplus cq'} + U_{qq'} b_{\oplus cq'}).
\end{eqnarray}
Of course, these operators have to fulfill the bosonic commutation relations:
\begin{eqnarray}
[a_{\oplus cq}, a^\dagger_{\oplus cq'}] &=& \delta_{qq'}, \label{appB bos comm rel}
\\
\left[a_{\oplus cq}, a_{\oplus cq'}\right] &=& [a^\dagger_{\oplus cq}, a^\dagger_{\oplus cq'}] = 0. \label{appB bos comm rel2}
\end{eqnarray}
Additionally, in order to ensure the diagonal form of $\hat{H}_{\mathrm{nd}}$ in terms of the new operators, the conditions
\begin{eqnarray}\label{appB diag cond}
[\hat{H}_{c}, a^\dagger_{\oplus cq} ] = \varepsilon_{cq} a^\dagger_{\oplus cq} &\Longleftrightarrow&  [\hat{H}_{c}, a_{\oplus cq} ] = -\varepsilon_{cq} a_{\oplus cq}\quad
\end{eqnarray}
have to be fulfilled. Note that also new energies $\varepsilon_{cq}$ have been introduced. The calculation of the transformation matrices $U$ and $V$ follows closely the guidelines in \cite{Avery76}. The result is
\begin{eqnarray}
U &=& \frac{1}{2} \left(\begin{array}{cc}  \sqrt{\frac{\varepsilon_{cq}}{2 \varepsilon_{0q}}} +\sqrt{\frac{\varepsilon_{0q}}{2 \varepsilon_{cq}}} & \quad- \sqrt{\frac{\varepsilon_{cq}}{2 \varepsilon_{0q}}} -\sqrt{\frac{\varepsilon_{0q}}{2 \varepsilon_{cq}}}
                                        \\           \sqrt{\frac{\varepsilon_{cq}}{2 \varepsilon_{0q}}} +\sqrt{\frac{\varepsilon_{0q}}{2 \varepsilon_{cq}}} &\quad \sqrt{\frac{\varepsilon_{cq}}{2 \varepsilon_{0q}}} +\sqrt{\frac{\varepsilon_{0q}}{2 \varepsilon_{cq}}}\;
                                        \end{array}\right)
\\
V &=& \frac{1}{2} \left(\begin{array}{cc}  \sqrt{\frac{\varepsilon_{cq}}{2 \varepsilon_{0q}}} -\sqrt{\frac{\varepsilon_{0q}}{2 \varepsilon_{cq}}} & \quad-  \sqrt{\frac{\varepsilon_{cq}}{2 \varepsilon_{0q}}} +\sqrt{\frac{\varepsilon_{0q}}{2 \varepsilon_{cq}}}
                                        \\           -\sqrt{\frac{\varepsilon_{cq}}{2 \varepsilon_{0q}}} +\sqrt{\frac{\varepsilon_{0q}}{2 \varepsilon_{cq}}} & \quad -\sqrt{\frac{\varepsilon_{cq}}{2 \varepsilon_{0q}}} +\sqrt{\frac{\varepsilon_{0q}}{2 \varepsilon_{cq}}}
                                        \end{array}\right), \nonumber
                    \\
\end{eqnarray}
where $\varepsilon_{0q}=\varepsilon_0|n_q|$ and
the transformed energies $\varepsilon_{cq}$ are given by
\begin{eqnarray}
\varepsilon_{cq} = \varepsilon_{0q}\sqrt{1+ 8W_{q,-q}/\varepsilon_{0}}. \label{transf en}
\end{eqnarray}
Using the abbreviations $\alpha_q = \frac{1}{2}\sqrt{\frac{\varepsilon_{0q}}{2 \varepsilon_{cq}}}$ and $\beta_q = \frac{1}{2}\sqrt{\frac{\varepsilon_{cq}}{2 \varepsilon_{0q}}}$ the new operators read, assuming $q>0$:
\begin{eqnarray*}
a^\dagger_{\oplus cq}  &=& (\alpha_q + \beta_q)  (b^\dagger_{\oplus c-q} + b^\dagger_{\oplus cq})
 \nonumber \\
 &+& (\beta_q-\alpha_q ) (b_{\oplus c-q} + b_{\oplus cq}),
\\
a_{\oplus cq}  &=& (\beta_q-\alpha_q  ) (b^\dagger_{\oplus c-q} + b^\dagger_{\oplus cq})
\nonumber \\
 &+&
(\alpha_q + \beta_q) (b_{\oplus c-q} + b_{\oplus cq}),
\\
a^\dagger_{\oplus c-q}  &=& (\alpha_q + \beta_q)    (b^\dagger_{\oplus c-q} - b^\dagger_{\oplus cq})
 \nonumber \\
 &+&
 (\alpha_q  -\beta_q) (b_{\oplus c-q} - b_{\oplus cq}),
\\
a_{\oplus c-q}  &=& (\alpha_q  -\beta_q) (b^\dagger_{\oplus c-q} - b^\dagger_{\oplus cq})
 \nonumber \\
 &+&
 (\alpha_q + \beta_q)    (b_{\oplus c-q} - b_{\oplus cq}).
\end{eqnarray*}
The inverse of the transformation is
\begin{eqnarray*}
b^\dagger_{\oplus cq}  &=& \frac{-(\alpha_q + \beta_q)  (a^\dagger_{\oplus c-q} - a^\dagger_{\oplus cq}) }{8 \alpha_q \beta_q}
\nonumber \\ &+&\frac{(\alpha_q  -\beta_q) (a_{\oplus c-q} + a_{\oplus cq}) }{8 \alpha_q \beta_q},
\\
b_{\oplus cq}  &=& \frac{(\alpha_q - \beta_q)   (a^\dagger_{\oplus c-q} + a^\dagger_{\oplus cq}) }{8 \alpha_q \beta_q}\nonumber \\ &+&
\frac{- (\alpha_q  +\beta_q) (a_{\oplus c-q} - a_{\oplus cq})}{8 \alpha_q \beta_q}
,
\\
b^\dagger_{\oplus c-q}  &=& \frac{(\alpha_q + \beta_q)  (a^\dagger_{\oplus c-q} + a^\dagger_{\oplus cq}) }{8 \alpha_q \beta_q}
\nonumber \\ &+&
\frac{ - (\alpha_q  -\beta_q) (a_{\oplus c-q} - a_{\oplus cq})}{8 \alpha_q \beta_q},
\\
b_{\oplus c-q}  &=& \frac{-(\alpha_q - \beta_q) (a^\dagger_{\oplus c-q} - a^\dagger_{\oplus cq}) }{8 \alpha_q \beta_q}\nonumber \\ &+&\frac{ (\alpha_q  +\beta_q) (a_{\oplus c-q} + a_{\oplus cq})}{8 \alpha_q \beta_q}.
\end{eqnarray*}
The inverse transformation can also be written in a more concise way:
\begin{eqnarray*}
b_{\tilde{F} jq} = \sum_{q'} \left( S_{\tilde{F}jqq'}a_{\tilde{F}jq'} + C_{\tilde{F}jqq'}a_{\tilde{F}jq}^\dagger \right), \label{appB trafo}
\end{eqnarray*}
where we have for the spin related coefficients $S$ and $C$:
\begin{eqnarray*}
S_{\oplus sqq'} &=& S_{\ominus sqq'} = S_{\ominus cqq'} = \delta_{qq'}, \label{appB Ss}
\\
C_{\oplus sqq'} &=& C_{\ominus sqq'} = C_{\ominus cqq'} = 0,
\end{eqnarray*}
while for the charge related ones, we have to take care of the sign of $q$. Let us choose $q>0$:
\begin{eqnarray*}
S_{\oplus cqq'} &=& \delta_{qq'} \frac{\alpha_q + \beta_q}{8\alpha_q \beta_q} - \delta_{q-q'} \frac{\alpha_q+ \beta_q}{8\alpha_q \beta_q},
\\
C_{\oplus cqq'} &=& \delta_{qq'} \frac{\alpha_q - \beta_q}{8\alpha_q \beta_q} + \delta_{q-q'} \frac{\alpha_q - \beta_q}{8\alpha_q \beta_q},
\\
S_{\oplus c-qq'} &=& \delta_{qq'} \frac{\alpha_q + \beta_q}{8\alpha_q \beta_q} + \delta_{q-q'} \frac{\alpha_q + \beta_q}{8\alpha_q \beta_q},
\\
C_{\oplus c-qq'} &=& \delta_{qq'} \frac{\alpha_q - \beta_q}{8\alpha_q \beta_q} - \delta_{q-q'} \frac{\alpha_q - \beta_q}{8\alpha_q \beta_q}. \label{appB Cq}
\end{eqnarray*}
The final result for the diagonalised Hamiltonian is then
\begin{eqnarray}
\hat{H}_\mathrm{ring} &=& \sum_{\tilde{F}j}\sum_{q \neq 0}  \varepsilon_{\tilde{F}jq} a_{\tilde{F}jq}^\dagger a^{\phd}_{\tilde{F}jq} + \varepsilon_0 \sum_{Fr \sigma} \frac{\mathcal{N}^2_{Fr\sigma}}{2} +\nonumber
\\
&+& \varepsilon_0 \tilde{\phi} \sum_{Fr\sigma} \mbox{sgn}(r) \mathcal{N}_{Fr\sigma} + \frac{1}{2} W_{00} \mathcal{N}^2,
\end{eqnarray}
where we introduced the energies
\begin{eqnarray}
\varepsilon_{\oplus sq}=\varepsilon_{\ominus sq}=\varepsilon_{\ominus cq}=\varepsilon_{0q}\mbox{ and }\varepsilon_{\oplus cq}=\varepsilon_{cq}.
\end{eqnarray}

\section{The bosonic correlator}\label{appB}
We evaluate in this appendix explicitly the bosonic part of the Green's function, Eq. (\ref{green-bosonic}),
 in the so called Luttinger limit,
where i) we assume a linear dispersion: $\varepsilon_{cq}=\varepsilon_{0q}/K_c$, with coupling constant
$K_c=\sqrt{1+8W_{q_0 -q_0}/\varepsilon_0}$;  ii) we assume an exponential decay of the parameter $C_{\oplus cqq}$ entering
the Bogoliubov transformation:  $C^2_{\oplus cqq}=C^2_{\oplus cq_0}e^{-q/q_c}$, with $q_c$ a momentum cut-off.
Let us then start from  Eq. (\ref{green-bosonic}):
 \begin{eqnarray}
\lefteqn{\left\langle  \psi^\dagger_{Fb\sigma}(x', -t) \psi^{\phd}_{Fb\sigma}(x, 0)\right\rangle_b=
\frac{1}{1-e^{-\alpha\frac{2\pi}{L}}}} \label{green-bosonic-app}\\
&\times&\exp\left\{
-\frac{1}{4}\sum_{q>0}\frac{e^{-\alpha q}}{n_q} \right.\left[ \sum_{\tilde F j}Q_{\tilde F j q}(t, {\rm sgn}(b)(x'-x))\right.
\nonumber\\
&+& \left.
2C^2_{\oplus cqq}[Q_{ c q}(t, (x'-x))+Q_{ c q}(t, -(x'-x))]
\right]\left.\right\},\nonumber
\end{eqnarray}
where we introduced the correlation function
\begin{eqnarray}
Q_{\gamma q}(t,x)&:=&i\sin \Bigl( \frac{\varepsilon_{\gamma q} t}{\hbar} -q x\Bigr)
 \\ &+&
\left[1-\cos\Bigl(\frac{\varepsilon_{\gamma q} t}{\hbar}-qx\Bigr)\right]
\coth\Bigl(\frac{\beta\varepsilon_{\gamma q} }{2}\Bigr).\nonumber
\end{eqnarray}
  In the low temperature regime we can approximate $\coth(\beta\varepsilon_{0q}/2)$ and $\coth(\beta\varepsilon_{cq}/2)$ by 1.
We perform now the Luttinger approximation and recall that $q=2n\pi/L$. Rewriting the sin and cos functions as exponentials and remembering the formula
\begin{equation}
\sum_{n>0}\frac{e^{an}}{n}=-\ln(1-e^a),\quad |e^a|<1,
\end{equation}
we arrive at
\begin{eqnarray}
\lefteqn{
\left\langle  \psi^\dagger_{Fb\sigma}(x', -t) \psi^{\phd}_{Fb\sigma}(x, 0)\right\rangle_b }\nonumber \\
&=&\left[1-z\Bigl(\alpha+\frac{1}{q_c},0\Bigr)\right]^{C^2_{\oplus cq_0}}\nonumber \\
&\times &\left[1-z\Bigl(\alpha, v_F t+{\rm sgn}(b)(x'-x)\Bigr)\right]^{-3/4} \nonumber \\
&\times & \left[1-z\Bigl(\alpha,{\rm v}_c t+{\rm sgn}(b)(x'-x)\Bigr)\right]^{-1/4}\nonumber \\
&\times&\left[1-z\Bigl(\alpha+\frac{1}{q_c},{\rm v}_ct+(x'-x)\Bigr)\right]^{-\frac{1}{2}C^2_{\oplus cq_0} }\nonumber \\
&\times&\left[1-z\Bigl(\alpha+\frac{1}{q_c},{\rm v}_ct-(x'-x)\Bigr)\right]^{-\frac{1}{2}C^2_{\oplus cq_0} },
\label{bosonic-2}
\end{eqnarray}
where we have defined the function
\begin{equation}
z(\alpha,x):=e^{(-\alpha+ix)\frac{2\pi}{L}}.
\end{equation}
Because the functions entering (\ref{bosonic-2}) are periodic in time, we can write the bosonic correlation function as a product of Fourier series. To this extent we use the relation
\begin{eqnarray}
(1-z)^\gamma = \sum_{k=0}^\infty \frac{\Gamma(k-\gamma)}{k!\Gamma (-\gamma)}z^k, \quad |z|<1,
\end{eqnarray}
which yields the final result
\begin{eqnarray}
\lefteqn{\hspace{-1cm}\left \langle \psi^\dagger_{F b \sigma} (x',-t) \psi_{F b \sigma}(x,0)\right\rangle_b = \left[ 1-z\left(\alpha + \frac{1}{q_c}, 0\right)\right]^{C^2_{\oplus cq_0}}} \nonumber
\\
&\times &\sum_{k=0}^\infty r_k\Biggl(\alpha, -\frac{3}{4}, \mbox{sgn}(b)(x'-x)\Biggr)e^{\frac{i}{\hbar}\varepsilon_0 kt} \nonumber \\ &\times&\sum_{l=0}^\infty s_{l}(b,x'-x)e^{\frac{i}{\hbar}\varepsilon_c l t}.
\label{appD bos}
\end{eqnarray}
Here we have defined
\begin{eqnarray*}
\lefteqn{r_k(\alpha, \gamma, x) := \frac{\Gamma(k-\gamma)}{k!\Gamma(-\gamma)}e^{k\frac{2\pi}{L}(-\alpha+ix)},
}\\
s_{l}(b,x'-x) &:=& \phantom{\times}\sum_{n_1=0}^{l} r_{n_1}\Biggl(\alpha, -\frac{1}{4},\mbox{sgn}(b)(x'-x) \Biggr)
\\
&& \hspace{-1.0cm}\times\sum_{n_2=0}^{l-n_1} r_{n_2}\left(\alpha + \frac{1}{q_c},-\frac{1}{2}C^2_{\oplus cq_0},(x'-x) \right)
\\
&&\hspace{-1.0cm}\times
\sum_{n_3 =0}^{l-n_1-n_2} r_{n_3}\left(\alpha + \frac{1}{q_c},-\frac{1}{2}C^2_{\oplus cq_0},-(x'-x) \right),
\end{eqnarray*}
where $\Gamma$ is the gamma function.

\end{document}